\documentclass[manuscript, screen]{acmart}
\pdfoutput=1
\usepackage{hyperref}
\usepackage{pdfpages}
\usepackage{booktabs}
\usepackage{graphicx}
\usepackage{lscape}
\usepackage{rotating}
\usepackage[utf8]{inputenc}
\usepackage{array}
\usepackage{ragged2e}
\usepackage{longtable}
\usepackage{adjustbox}
\usepackage{rotating}
\usepackage[longtable]{multirow}
\usepackage{titlesec}
\usepackage{hyperref}
\usepackage{float}
\restylefloat{table}
\usepackage{pifont}
\newcommand{\cmark}{\ding{51}}%
\newcommand{\xmark}{\ding{55}}%

\AtBeginDocument{%
  \providecommand\BibTeX{{%
    \normalfont B\kern-0.5em{\scshape i\kern-0.25em b}\kern-0.8em\TeX}}}

\setcopyright{acmcopyright}
\copyrightyear{2022}
\acmYear{2022}
\acmDOI{XXXXXXX.XXXXXXX}

\acmJournal{JACM}
\acmVolume{37}
\acmNumber{4}
\acmArticle{}
\acmMonth{22}

\acmPrice{15.00}
\acmISBN{978-1-4503-XXXX-X/18/06}

\begin{document}

\title{Private Knowledge Sharing in Distributed Learning: A Survey }

\author{Yasas Supeksala}
\affiliation{%
 \institution{Swinburne University of Technology}
 \streetaddress{Jhon St}
 \city{Melbourne}
 \state{Victoria}
 \country{Australia}}

\author{Dinh C. Nguyen}
\affiliation{%
  \institution{Purdue University}
  \city{West Lafayette}
  \state{Indiana}
  \country{USA}}

\author{Ming Ding}
\affiliation{%
  \institution{DATA61-CSIRO}
  \streetaddress{Eveleigh}
  \city{Sydney}
  \state{NSW}
  \country{Australia}}

\author{Thilina Ranbaduge}
\affiliation{%
  \institution{DATA61-CSIRO}
  \streetaddress{Eveleigh}
  \city{Sydney}
  \state{NSW}
  \country{Australia}}
  
\author{Calson Chua}
\affiliation{%
 \institution{Swinburne University of Technology}
 \streetaddress{Jhon St}
 \city{Melbourne}
 \state{Victoria}
 \country{Australia}}
 
\author{Jun Zhang}
\affiliation{%
 \institution{Swinburne University of Technology}
 \streetaddress{Jhon St}
 \city{Melbourne}
 \state{Victoria}
 \country{Australia}}

 \author{Jun Li}
\affiliation{%
  \institution{Nanjing University of Science and Technology}
  \streetaddress{Nanjing}
  \country{China}}

 \author{H. Vincent Poor}
\affiliation{%
  \institution{Princeton University}
  \streetaddress{Princeton}
  \country{USA}}

\begin{abstract}
The rise of Artificial Intelligence (AI) has revolutionized numerous industries and transformed the way society operates. Its widespread use has led to the distribution of AI and its underlying data across many intelligent systems. In this light, it is crucial to utilize information in learning processes that are either distributed or owned by different entities. As a result, modern data-driven services have been developed to integrate distributed knowledge entities into their outcomes. In line with this goal, the latest AI models are frequently trained in a decentralized manner. Distributed learning involves multiple entities working together to make collective predictions and decisions. However, this collaboration can also bring about security vulnerabilities and challenges. This paper provides an in-depth survey on private knowledge sharing in distributed learning, examining various knowledge components utilized in leading distributed learning architectures. Our analysis sheds light on the most critical vulnerabilities that may arise when using these components in a distributed setting. We further identify and examine defensive strategies for preserving the privacy of these knowledge components and preventing malicious parties from manipulating or accessing the knowledge information. Finally, we highlight several key limitations of knowledge sharing in distributed learning and explore potential avenues for future research.

\end{abstract}

\maketitle
\section{INTRODUCTION}
\label{sec:intro}
 
Owing to the amplified sensitivity and enhanced computing capabilities of end-user applications and Internet of Things (IoT) devices, an enormous amount of data is being generated and collected at an unprecedented pace. The traditional machine 
learning (ML) systems that integrate such data into a centralized device to process are 
becoming less feasible in real-world scenarios because of communication constraints  and the computational overhead of the 
large data silos. As a promising solution, \emph{distributed learning} has garnered substantial attention in contemporary machine learning applications~\cite{verbraeken2020survey}. Distributed learning was introduced as a mechanism to boost the efficiency, accuracy, and data interpretation capabilities of distributed end nodes responsible for data gathering and processing. By leveraging distributed learning, users can obtain more precise predictions with significantly lower computational overhead, rendering distributed learning more applicable to applications that involve distributed data~\cite{peteiro2013survey}. 

Distributed learning has gained widespread adoption in industrial applications~\cite{nassef2022survey}. However, for each end node (i.e., data provider) in a distributed learning framework, it is essential to enhance its processing power to enable distributed learning. This requires transitioning from single-threaded algorithms to parallel algorithms~\cite{leighton2014introduction}. Distributed architectures for machine learning can be classified into four primary frameworks, depending on the approach to parallelization: supervised, unsupervised, semi-supervised, and deep reinforced learning. In supervised machine learning, labeled data is used to train ML models. When this concept is extended to a distributed setting, where labels are stored in distributed nodes, it becomes distributed supervised learning.
Similarly, in unsupervised learning, the training process can be described as distributed unsupervised learning in an ML model that generates global predictions using unlabeled data from distributed nodes. Semi-supervised learning combines both supervised and unsupervised mechanisms and can train an ML model using a small amount of labeled data while generating predictions for a large volume of unlabeled data.
Reinforcement learning is an ML mechanism that rewards desired behaviors and penalizes undesired ones. When reinforcement learning is accomplished through experience divided between actors and learners, it is referred to as distributed reinforcement learning~\cite {kapturowski2018recurrent}. 
Reinforcement learning, in essence, can be considered an autonomous deep learning architecture, as it does not fall within the purview of any supervised, unsupervised, or semi-supervised mechanisms. Instead, it utilizes environmental cues to make predictions and improve performance through trial and error ~\cite{asad2021federated}. \par

It's important to note that federated learning (FL) architectures, which have been introduced more recently, cannot be straightforwardly categorized as distributed learning mechanisms due to the independent training characteristics and network federation of participants. However, FL plays a significant role in current collaborative learning mechanisms. The difference between FL and distributed learning lies in how these techniques achieve collaborative learning.

In distributed learning, local devices build their models based on their local datasets and then collaborate with other distributed parties or the global server to update the model~\cite{liu2022distributed}. However, in federated learning, the distributed parties train a model entirely, and the \emph{knowledge of the model} is shared through an aggregated model with other participants. In a generic distributed learning setting, local clients can share various types of information with each other, such as certain processed forms of data, models, features, predictions, and so on. On the other hand, in FL, local clients share ML models with each other.\par
       
When examining these distributed learning mechanisms, it is crucial to understand how distributed learning is accomplished. In any reliable distributed learning framework, a robust communication mechanism should be established between the participating nodes in the network, system, or lattice to achieve distributed learning~\cite{chen2021distributed}. Therefore, communication plays a critical role in a distributed learning architecture, facilitating interactivity and collaboration between nodes. To achieve a more robust and reliable distributed learning model, the participants should complete at least tens of communication rounds, which also depend on the computation capability of the learning platform.

While the communication stage of a distributed learning architecture is critical, the approach to achieving communication will differ from one distributed model to another, depending on the architecture and application~\cite{balcan2012distributed}. However, the main components shared in each communication round are typically sample data batches, model outputs, and, in some cases, both. This paper aims to differentiate the various components that are common to deep learning models. In doing so, any parameter associated with the model's output or any residual element that remains after the training phase is considered the \emph{knowledge} possessed by the model.\par

The security of communication rounds in distributed learning is a significant concern, as the knowledge components exchanged between nodes can consist of vulnerabilities in any distributed learning architecture~\cite{bouacida2021vulnerabilities}. Such communication channels between nodes provide a larger attack surface for adversaries, making it imperative to secure these channels. Increasing the number of communication rounds required to train an ML model also exposes the learning system to various vulnerabilities.

While sharing data samples among model participants can create vulnerabilities, these can be mitigated using privacy techniques based on homomorphic encryption \cite{froelicher2021scalable} or differential privacy (DP) mechanisms~\cite{xie2021differential}. Unfortunately, there are not many robust defense frameworks proposed in the current cybersecurity research landscape to address the security concerns in modern distributed deep learning architectures~\cite{liu2022distributed}.

\begin{figure}[t!]
    \centering 
    \includegraphics[width=0.75\linewidth]{ {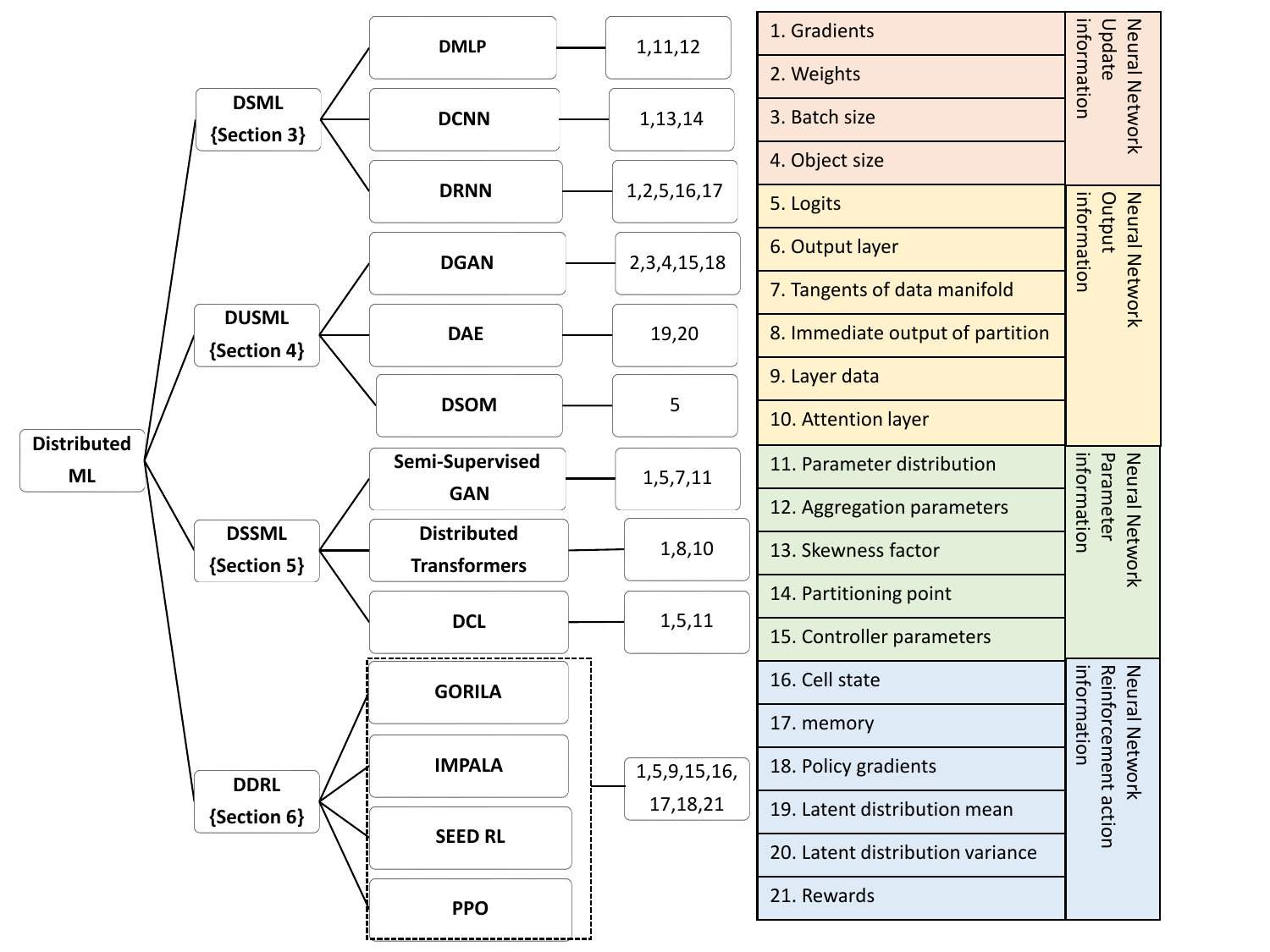} }
    \caption{Taxonomy of Distributed Learning Models: The numbered items in each box represent the \emph{knowledge components} we discuss through the paper.} 
	\label{Fig:Taxonomy}
	\vspace{-1mm}
\end{figure}

\subsection{Comparison and Contribution} 

In recent years, there has been a growing interest in deep learning settings, including research that explores the characteristics and applications of different deep learning paradigms, such as those used in edge computing and wireless networks, as well as the data privacy of distributed entities. These studies mainly focus on the implications of distributed systems compared to traditional centralized systems and highlight their limitations and challenges~\cite{verbraeken2020survey,hu2021distributed,sams2020ddrl,Filho2022dmline,nguyen2021federated}.

However, there is limited research that comprehensively reviews the \emph{knowledge component} of deep learning models in distributed learning architectures and the potential vulnerabilities that may arise when these components are used in a distributed setting. This paper makes a significant contribution to the field of distributed deep learning architectures by addressing the gap in knowledge regarding the potential vulnerabilities that arise in this context. By examining these vulnerabilities and discussing defensive strategies that can be employed to protect the privacy of knowledge components and prevent them from being manipulated or accessed by malicious entities, this research advances the ongoing efforts to ensure the security and privacy of distributed deep learning architectures.

This research builds on the existing body of work in the field of distributed systems and their relation to conventional systems, including the survey conducted by \cite{verbraeken2020survey}. While these studies provide valuable insights into the transition from conventional to distributed systems, this research focuses specifically on the privacy implications of using distributed deep learning architectures, which have become increasingly vital in modern machine learning applications. By doing so, this research aims to provide a more comprehensive understanding of the privacy and security challenges associated with distributed deep learning architectures and to inform the development of effective defensive strategies to mitigate these challenges.
While prior research, such as Sams et al.~\cite{sams2020ddrl}, has focused on the implementation of distributed learning in specific architectures, answering questions such as how reinforcement learning can be achieved in a distributed environment and other works, such as \cite{Filho2022dmline} and \cite{duc2019machine}, have examined ways to improve communication efficiency, but none have comprehensively addressed the vulnerabilities associated with knowledge sharing in distributed deep learning architectures, and how these vulnerabilities can be mitigated using privacy-enhancing techniques.

In contrast to previous research, our study provides a comprehensive examination of various distributed deep learning architectures. We identify the specific parameters that are shared during communication rounds across these architectures, with a focus on the knowledge components and the vulnerabilities associated with them. Furthermore, we evaluate the effectiveness of existing defense mechanisms against these identified vulnerabilities. Table~\ref{tab:summary_dl_papers} summarizes our contributions in comparison to other surveys presented in the literature. Our key contributions are as follows:
 

\begin{enumerate}
    \item   We conduct an extensive investigation of the fundamental distributed learning architectures
and categorize their learning strategies and respective 
\emph{knowledge components}.
\item  We identify and discuss critical vulnerabilities associated with 
each knowledge component, assuming the communication phase in distributed learning is prone
to attacks.
\item We explore different attack implementations and methodologies in the
recent literature that can exploit different knowledge components. 
\item Finally, we outline robust defense mechanisms that
can be integrated into distributed systems to overcome those vulnerabilities that 
have been identified in distributed learning systems and how they can contribute to the realization of user-friendly, privacy-preserved AI systems.
\end{enumerate}

\subsection{Structure of the Survey}    
The paper is structured as follows. We begin with an introduction to knowledge sharing in distributed learning architectures in Section \ref{sec:knowledge_sharing_in_dl}. In Section \ref{sec:knowledge_sharing_in_sdl}, we provide a detailed analysis of knowledge sharing in distributed supervised learning, including an examination of the knowledge components, potential attacks, and defensive strategies. We then discuss knowledge sharing in distributed unsupervised, semi-supervised, and reinforcement learning in Sections \ref{sec:knowledge_sharing_in_udl}, \ref{sec:knowledge_dssl}, and \ref{sec:know_DDRL}, respectively. In Section \ref{sec:limitations_and_future_directions}, we discuss important limitations of knowledge sharing in distributed learning architectures and suggest several future research directions. Finally, in Section \ref{sec:conclusion}, we conclude the paper. The flow of the article is illustrated in Figure \ref{Fig:Taxonomy}. By organizing our discussion in this manner, we provide a clear and structured overview of knowledge sharing in distributed learning architectures, enabling readers to gain a comprehensive understanding of this complex and important topic.

\begin{table}
\footnotesize
\begin{longtable}[h!]{|m{0.04\linewidth}|m{0.04\linewidth}|m{0.04\linewidth}|m{0.04\linewidth}|m{0.04\linewidth}|m{0.04\linewidth}|m{0.26\linewidth}|m{0.27\linewidth}|} 
\caption{Summary of related works and our new contributions. Legend: \cmark signifies if discussed components are present or addressed while \xmark signified the absence of components, (DSL (Distributed supervised Learning), DUSL (Distributed Unsupervised Learning), DSSL (Distributed Semi-Supervised Learning), DDRL (Distributed Deep Reinforcement Learning)).}
\label{tab:summary_dl_papers} \\\hline

\multirow{2}{*}{Paper} &
\multirow{2}{*}{\begin{tabular}[c]{@{}l@{}}Att.\\\&\\Def.\end{tabular}} & 
\multicolumn{4}{c|}{Learning architecture} &
\multirow{2}{*}{\begin{tabular}[l]{l@{}c@{}}Discussion on \\\ knowledge sharing\end{tabular}}  &
\multirow{2}{*}{Shared knowledge} \\* \cline{3-6}

&  & 
\begin{sideways}DSL\end{sideways} &
\begin{sideways}DUSL\end{sideways} &
\begin{sideways}DSSL\end{sideways} & 
\begin{sideways}DDRL\end{sideways} &
&
 \endfirsthead \hline

~\cite{verbraeken2020survey} & \xmark & \cmark & \cmark  & \cmark & \cmark & Touches knowledge sharing on the basis of the communication phase. & 
A discussion of the implications of distributed systems over conventional machine learning systems with challenges and limitations. \\ \hline

~\cite{hu2021distributed} & \cmark & \xmark & \xmark & \xmark & \cmark & 
Constructs a discussion about knowledge sharing based on increasing communication efficiency and model convergence. & 
A comprehensive survey on the application of distributed learning considering trade-offs in communication networks. \\ \hline

~\cite{sams2020ddrl} & \cmark & \xmark & \xmark  & \xmark & \cmark & 
The authors clearly demonstrated how model participants share the knowledge. & 
Survey on strategies that adapt distributed learning to edge and fog computing. \\ \hline

~\cite{Filho2022dmline} & \xmark & \cmark & \xmark & \xmark & \xmark &
Discusses different approaches that can facilitate edge intelligence but does enumerate the knowledge \par{} of the AI model. & 
An investigation of challenges of running ML or Deep Learning models at the edge of the network in a distributed manner. \\ \hline

~\cite{zerka2020systematic} & \cmark & \cmark &\cmark & \xmark & \cmark &
Discusses the implementations of classification algorithms in distributed setting but do not concentrate on the specific knowledge components. &
A survey that addresses the privacy concerns of patient data that occur due to data centralization and alternative approaches to facilitate distributed of data processing. \\ \hline

~\cite{sapio2021scaling} & \xmark & \cmark & \xmark & \xmark & \xmark & Demonstrates how the aggregation of machine learning algorithm initiated in a distributed learning setting but do not focus on specific ML architecture. & 
Investigates ways of aggregating data in distributed machine learning systems. \\ \hline

~\cite{duc2019machine} & \cmark & \cmark & \cmark & \xmark & \xmark & Looks into knowledge sharing only in the aspect of infrastructure optimization. &
A Survey on resource-aware device placement in a distributed edge network~using the state of the art knowledge of distributed learning. \\ \hline

~\cite{nguyen2022federated} & \cmark & \cmark & \cmark & \cmark & \cmark & 
Discusses how the communication is being initiated in different FL architectures but do not elaborate on the knowledge components. & 
Identify and investigate how federated learning can be adopted into the healthcare. \\ \hline

Our Work & \cmark & \cmark & \cmark & \cmark & \cmark & 
Investigates state-of-the-art distributed machine architectures and brings out the knowledge components that are being shared. & 
Identifies the different knowledge components associated with each distributed learning architecture and illustrates their vulnerabilities. Then we identify current exploitations on those vulnerabilities and present privacy preservation techniques. \\ \hline
\end{longtable}
\end{table}

\section{KNOWLEDGE SHARING IN A DISTRIBUTED LEARNING SETTING}
\label{sec:knowledge_sharing_in_dl}

\begin{figure}[t!]
    \centering 
    \includegraphics[width=1.0\linewidth]{ {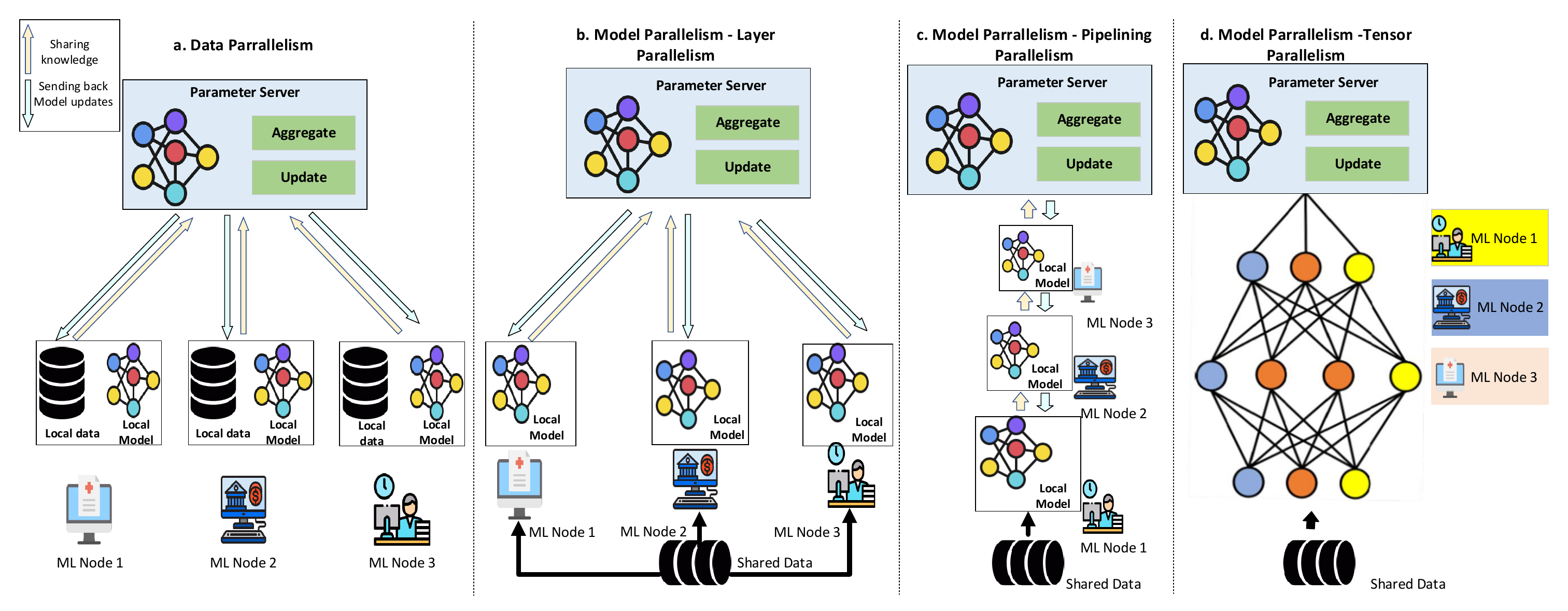} }
    \caption{Distributed learning: part "a" illustrates distributed learning achieved through data parallelism while parts "b" to "d" demonstrate various architectures implemented on the fundamental idea of model parallelism.} 
	\label{fig:overview_DL}
\end{figure}

Distributed learning is a powerful technique that combines multiple nodes' computational resources and data to create high accuracy. These high-performance machine learning models can handle vast amounts of data~\cite{verbraeken2020survey}.  By leveraging the collective power of multiple participants, distributed learning allows for sophisticated decision-making in today's intelligent systems, which are increasingly reliant on artificial intelligence.
When examining various distributed learning architectures, it is apparent that distributed learning can be considered a form of parallel learning, where traditional single-threaded algorithms are transformed into parallel.  
The method used to achieve this parallelization is a vital aspect when defining the distributed learning strategy~\cite{hegde2016parallel}, which will be thoroughly explored in depth in this paper. Distributed learning parallelism has two main approaches: (1) \emph{data parallelism} and (2) \emph{model parallelism}.  In this paper, we do not dive deep into the distribution
achieved by data parallelism, as shown in Figure \ref{fig:overview_DL}. While data parallelism has been well-studied in the current research landscape and privacy concerns related to privacy-preserving data applications have been primarily addressed, and this paper will focus on model parallelism, which is concerned with how knowledge is generated through machine learning models. This aspect of distributed learning is crucial as an ML model can learn and identify unique patterns and make informed decisions based on this knowledge.

The deep learning model components can be classified as AI knowledge due to the adroitness acquired from prior experience. When looking into 
model parallelism, it can be achieved via several methodologies. Although in modern AI architectures, most of the time, some subset of data samples might also be integrated with the distributed training instances to increase the training accuracy, the impact of such integration is negligible when looking into the role of local model contribution. Hence, 
our survey mainly concentrates on identifying the knowledge components explicitly associated with the model parallelism phase. Figure~\ref{fig:overview_DL} illustrates how model parallelism and data parallelism facilitate distributed
training and how AI's knowledge is associated with the model. From Figure~\ref{fig:overview_DL}, it can clearly be observed how distributed learning achieves parallelisms in the model and data. 

Furthermore, model parallelization can be classified into three main architectures.~\cite{10.1145/3442442.3452055}.
\begin{itemize}
    \item \emph{Tensor parallelism}: Tensor parallelism is the most 
    widely used model parallelism technique. In Tensor parallelism, model weights, 
    gradients, and optimizer states are split across devices. Furthermore, it splits
    individual weights using distributed computations of the model's specific 
    operations, modules, or layers. When tensor parallelism is used on the model, 
    its forward and backward propagation is distributed.
    
    \item \emph{Layer-wise parallelism}: Also known as optimizer state sharing, 
    this method can share a single replica of the optimizer across data-parallel 
    ranks with no redundancy across devices.

    \item \emph{Layer pipelining}: Layer pipelining is capable of keeping individual weights intact but partitions the set of weights. When larger models need to be 
    partitioned spatially, layer pipelining can be used, bringing the hardware
    utilization to make the hardware utilization more efficient and effective. Moreover, for pipelining parallelism, there are 
    two main approaches, known as synchronous pipeline parallelism and asynchronous
    pipeline parallelism. When looking into these methods, those differences are 
    in the context of how the mechanisms achieve gradient synchronization among
    the adjacent training iterations.
\end{itemize}

In the following Sections \ref{sec:knowledge_sharing_in_sdl}-\ref{sec:know_DDRL}, we will investigate the distributed deep learning 
architectures that achieve distributed learning using the model parallelism mentioned above. 
We identify the critical shared information, which can be identified as \emph{knowledge}. 
Figure~\ref{fig:overview_DL} demonstrates how 
traditional ML architectures achieve deep learning using the above-mentioned 
parallelisms.
Due to the extensive number of working schemes, we have identified the \emph{sharing points} or \emph{knowledge components} through these architectures. Considering that we have found more than twenty of these knowledge components, as demonstrated in Figure \ref{Fig:Taxonomy}, we categorize these knowledge components into four main categories for better management of the 20 plus knowledge components we are discussing. This categorization will allow us to make informed decisions by grouping and comparing the components based on their role and characteristics in a distributed learning environment.
\begin{itemize}
    \item \emph{Neural Network Update information}: The distributed entities in the system learn through a series of iterations. The distributed entities and/or the global model are updated with the results during each iteration. We refer to these shared points as a \emph{knowledge category}. Our investigation has identified \emph{gradients} and \emph{weights} as the primary components of this knowledge category that contribute to updates in a neural network. However, for more complex structures, we have also found that the data's \emph{batch size} and \emph{object size} can also be grouped under this knowledge category.
    \item \emph{Neural Network Output information}: Distributed learning is the collaboration of different entities to achieve optimized status prediction~\cite{verbraeken2020survey}. The specific methods of collaboration depend on the ML architecture and implementation. However, there are two main types of collaboration that can be identified: collaboration during training and collaboration after training, which involves sharing the output of a training iteration. We refer to this shared output information as \emph{neural network output information}. Our investigation into knowledge components has identified \emph{logits}, \emph{layer data including attention and output layers}, and \emph{immediate output of partition} as the output information that is shared to facilitate distributed learning.
    \item \emph{Neural Network Parameter information}: While we do not consider \emph{parameters} as a knowledge component, we consider the distribution of parameters as a knowledge component. Thus, any parameter that determines a course of action is considered a knowledge type that falls under the category of \emph{neural network parameter information}. Based on this, we have identified \emph{parameter distribution}, \emph{aggregation parameters}, \emph{skewness factor}, \emph{tangents of data manifold}, \emph{partitioning points}, and \emph{control parameters} as the knowledge components that can be classified as part of \emph{neural network parameter information}.
    \item \emph{Neural Network Reinforcement action information}: Deep reinforcement learning typically involves complex and computationally intensive architectures. When these are extended into a distributed setting, the complexity increases further. In the implementation of these complex reinforcement mechanisms, there are dedicated knowledge components that assist in the reinforcement process. We have identified the components of \emph{cell state}, \emph{memory}, \emph{latent distribution mean and variance}, \emph{policy gradients}, and \emph{rewards} and classified them under the knowledge category of \emph{neural network reinforcement action information}.
\end{itemize}
In the following sections, we will identify the various knowledge components in different learning architectures and analyze the vulnerabilities to different attacks on the knowledge generated by AI models. We will also evaluate the defensive mechanisms against such attacks.

\section{KNOWLEDGE SHARING IN DISTRIBUTED SUPERVISED LEANING}
\label{sec:knowledge_sharing_in_sdl}
 
Supervised learning plays a significant role when considering 
\emph{Knowledge sharing among AI}. The use of supervised learning is prevalent in modern AI applications, and this approach can also be implemented in a distributed setting. To gain a deeper understanding of the knowledge elements involved in the collaborative learning stage, we have broken down distributed supervised learning into three main architectural types. We will delve into how these traditional machine learning architectures are utilized in deep learning and examine the knowledge components associated with each one in the following sections.

\subsection{Distributed Training for Multi-Layer Perceptrons}
\label{sec:dsl_mlp}

A Multi-Layer Perceptron (MLP) is a classic supervised learning method composed of several layers: An input layer, a hidden layer, and an output layer. The input layer receives the input data to be processed, the output layer carries out a specified classification or prediction task, and the hidden layers are located between the input and output layers, functioning as the computational engine for the MLP.~\cite{gardner1998artificial}.
In the broader field of AI, neural network architectures of the traditional type are commonly referred to as MLP. These architectures have the ability to learn the representation of data used for training and then make predictions or carry out classification tasks. 

The distributed MLP architecture is a method for implementing a traditional MLP neural network using multiple devices or machines. Utilizing the computational power and memory of multiple machines allows for the training of larger and more complex models than can be trained on a single machine.
In this architecture, the training data is split among the machines, with each machine training a separate version of the model on its assigned portion of data. These versions of the model are then consolidated to form the final model. The typical approach in the industry is to assign different layers to different components of the learning requirements. And then, in the inference stage, MLPs are collaboratively trained to produce the output~\cite{gardner1998artificial}. In the following section, we will present the latest developments in distributed MLP implementations and detail the various knowledge components that can be obtained from them.

\subsubsection{The Knowledge Components}\hfill\\
Chang et al.~\cite{chang2018deeplinq} proposed a novel ML model called 
\emph{DeepLinQ}, an efficient supervised distributed architecture. The proposed model
achieves efficiency by accelerating the training of deep learning models through model compression, distributed optimization, and hardware-aware deep learning. The authors use a blockchain mechanism~\cite{chang2018deeplinq} as a privacy mechanism, and the general distributed learning architecture is constructed on MLP. When
looking into how they are achieving this collaborative learning, there are five primary
design considerations that can be taken into account. These design considerations can be identified as \emph{shared ledgers among silo systems}, \emph{storage}, \emph{smart contracts}, \emph{data anonymity}, and \emph{multi-layer blockchain}. We mainly consider the design consideration of \emph{shared ledgers among silo systems} since it is responsible for knowledge sharing in this architecture.

The authors have implemented distributed learning using prioritized gradient descent, which updates the global model using the parameters of the data points with the highest contribution. It is important to consider both model acceleration and convergence mechanisms when relating the knowledge discussed in ~\cite{chang2018deeplinq} to \emph{knowledge sharing}. When examining the model convergence used in their architecture, it can be seen that the model participants will distribute the data points and their corresponding calculations of priority and gradient at the first stage. After the training stage, the updated gradients are reported to a parameter server that maintains the latest model parameters. This training process continues until satisfactory performance is achieved.

\begin{figure}[t!]
    \centering 
    \includegraphics[width=0.5\linewidth]{ {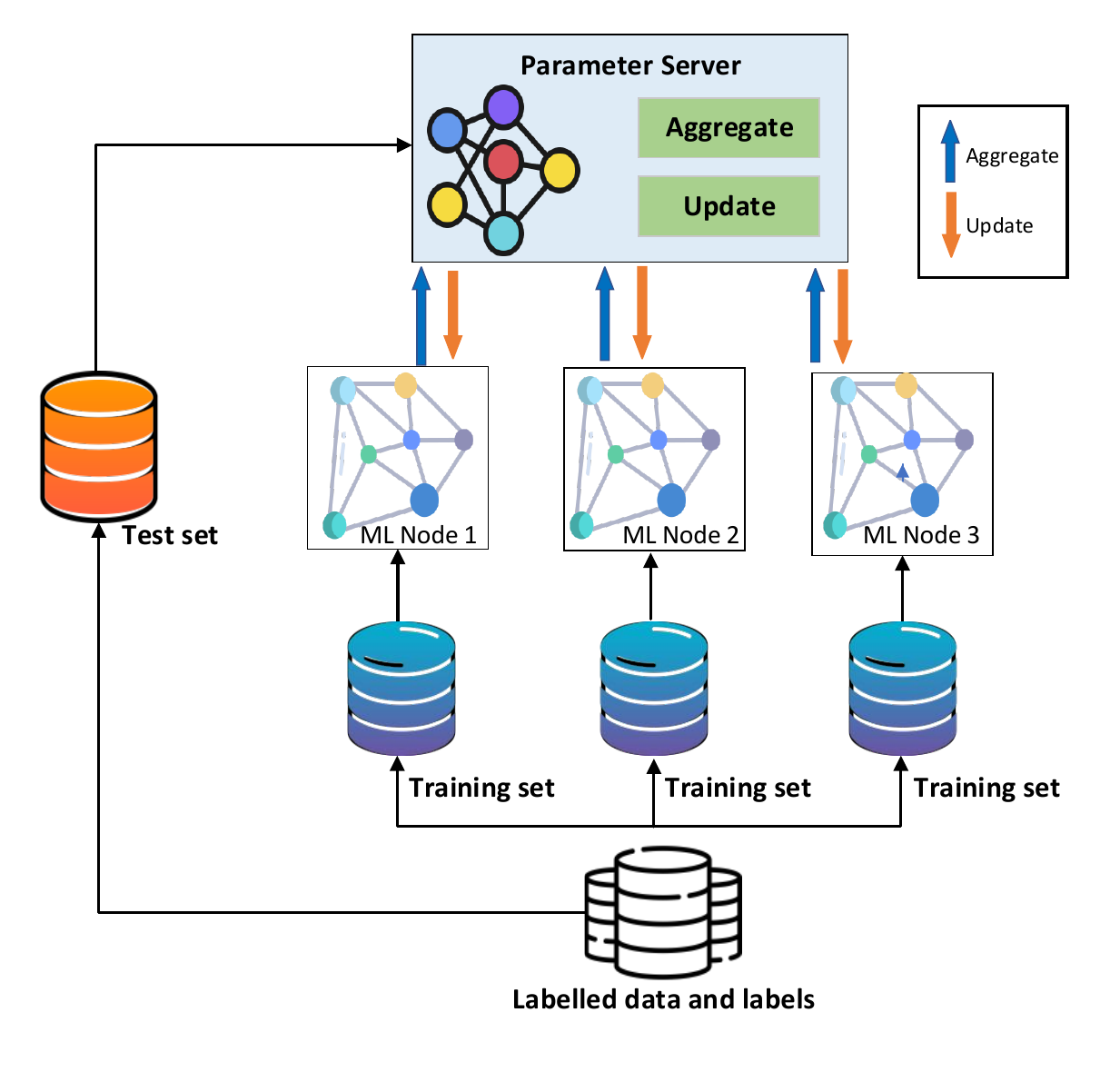} }
    \caption{Knowledge sharing in supervised learning} 
	\label{fig:knowledge_sharing_in_DSL}
\end{figure}

Figure~\ref{fig:knowledge_sharing_in_DSL} demonstrates this architecture briefly. When 
looking into how these models collaboratively train and send model parameters to sever, it can 
be noted that \emph{knowledge} sharing is happening between the model participants. As illustrated in
Figure~\ref{fig:knowledge_sharing_in_DSL}, the knowledge that is communicated 
between the participants can be directly identified as gradients. This is because, during the learning process, the global model updates its local models by directly utilizing the sent gradients, and the gradients are shared with all participants.

Based on these observations, we identify \emph{gradients} are a significant
\emph{knowledge component} that can be used to facilitate the learning
process of distributed MLPs.
The authors of~\cite{xia2019rethinking} are using MLP to preserve the 
privacy of the transport layer and make it more effective in distributed machine 
learning systems. The authors also pointed out that the current distributed MLP implementations have a significant delay due to their synchronous nature. This means that in these systems, the model participants must update their parameters before progressing to the next training iteration.

In order to address these issues, Xia et al. ~\cite{xia2019rethinking} suggest 
proposed an alternative mechanism that can continue training despite packet loss in communications. When looking into the 
knowledge-sharing scheme that is integrated into this mechanism, we can see that the knowledge-sharing is conducted through \emph{parameter distribution}
and \emph{aggregation parameters}. So unlike a method that completely shares all gradients
as the \emph{knowledge}, this proposed mechanism mainly relies on parameter 
distribution and data aggregation parameters to contribute to the main learning 
architecture. Thus, we consider \emph{parameter distribution} and \emph{aggregators}
also as knowledge components that persist in distributed MLP systems. 

 It should be brought forward that, in our investigation, the model
parameters would not be considered as knowledge since parameters are not an output of 
an experience but a component that facilitates architectural integrity. However,
the \emph{parameter distribution} is an outcome of one or multiple learning iterations. Thus,
it can be considered knowledge. 

When looking into the \emph{aggregators}, having the ability to perform parameter aggregation, or determining how to combine multiple parameters during the inference phase, can be seen as a valuable skill. This is because adjusting the parameter aggregation strategy can significantly impact the training accuracy of the model. The ability to quickly alter the aggregation parameters allows for flexibility in fine-tuning the model's performance ~\cite{guerraoui2018hidden}. This paper focuses on identifying the knowledge components that make up a system and then examining how these components can be protected in a manner that preserves privacy.  In Section~\ref{sec:att_dsl}, we discuss the different 
vulnerabilities associated with each of these knowledge components and illustrate how 
different attacks can manipulate these knowledge components in distributed MLP systems.

\subsection{Distributed Convolutions Neural Networks}
\label{sec:DCCNN}
Convolutional Neural Networks (CNNs) are a type of artificial neural network designed for image recognition and processing~\cite{ANIRUDH2021151}. They are made up of three main layers: the convolution layer, the pooling layer, and the fully connected layer. CNNs work by analyzing the patterns in images through the use of a grid-like structure that represents the image in binary form. These models are widely used in a variety of image-related tasks, such as object identification and classification. \textcolor{black}{ Each of these layers contributes towards a specific role in the decision-making process and can be remotely aggregated in a distributed setting. However, when elaborating on each layer, it should be noted that only Fully-connected layer and the convolution layer contributed to the model parameters, while the non-linearity layer and the pooling layer contributed to the generalization or saturation of the output and down-sampling for the optimizations accordingly.}\par
The convolution layer plays a central role in the architecture of a CNN. It consists of a set of filters, also known as kernels, which are used to identify specific patterns in an input image. These filters are smaller in size than the input image and slide over the image, performing a mathematical operation called convolution to produce an activation map. The parameters of the filters are learned through the training process. The convolution layer is essential for extracting features from the input image and passing them on to the successive layers of the network~\cite{ANIRUDH2021151}. 

Distributed Convolutional Neural Networks are a variant of CNNs that are designed to operate in a distributed setting, where multiple participants work together to perform image recognition and processing tasks. In a distributed CNN, knowledge is distributed among the participants through layer pipelining, a process in which each participant is responsible for processing a specific layer of the network. There are several variations of distributed CNNs, each with its unique set of knowledge components that are shared among the participants. In the following section, we examine these variations and discuss the knowledge components that are shared in a distributed CNN setting.

\subsubsection{The Knowledge Components} \hfill\\
In our examination of the knowledge elements related to distributed CNNs, we delve into various implementations of distributed CNNs.
Boulila et al.~\cite{boulila2021rs} proposed a distributed CNN that processes 
satellite images in two steps. The first step concerns ingesting big data, which 
splits data across distributed nodes. Instead of using the traditional 
size-based approach to split the images, the authors have introduced a new 
parallel method for data preparation and partition by  splitting large-scale image data 
into small portions and then applying a supervised classification algorithm 
based on \emph{Maximum Likelihood} accordingly to prepare the training data set. 
The second step of their approach aims to apply distributed CNN to classify 
the images. 

Stahl et al.~\cite{stahl2021deeperthings} proposed a mechanism in which the 
fully connected layer and convolution layers (feature and weight-intensive)
are partitioned to combine with a communication-aware layer fusion method. This
partitioning method enables holistic optimization across the layers.
Also based on the general layer pipelining mechanism, Zhang et al.~\cite{zhang2020adaptive} 
proposed a distributed CNN mechanism where it obtains the distributed training \textcolor{black}{by dividing the convolutional layers into numerous small layers, separate computational tasks, and then these tasks are assigned to edge nodes dynamically based on their current operational status}.
This mechanism, known as \emph{ADCNN} ~\cite{zhang2020adaptive}, offers an efficient distribution among a set of edge devices.
The main layers used for this distribution are the convolution layer and the 
pooling layer. However, the knowledge component that has been used in the 
described CNN architectures is mainly the \emph{convolution and pooling layers},
while the fully connected layer is placed in the global model. 

Even when knowledge is distributed among the participants based on different layers, the fully connected layer cannot be entirely ignored. This is because, in certain cases, the first and second parts of the fully connected layer may be shared among the participants during collaborative training. Therefore, it is important to consider the fully connected layer when examining the knowledge that is shared in a distributed CNN model~\cite{stahl2021deeperthings}. We also
identified that to obtain more optimized and accurate output, in some work  \cite{park2019accelerated}, the authors have used resource-aware layer placements to facilitate resource awareness.
The approaches proposed in \cite{stahl2021deeperthings} and \cite{park2019accelerated} 
have used \emph{skewness factor}, \emph{demand layer}, and \emph{partitioning point}
as the knowledge generated by their corresponding training models.  However, the 
demand layer can be taken as a component of the output layers. Thus, we considered \emph{output layers}, the \emph{skewness factor}, and the \emph{partitioning points}
as knowledge components under distributed CNN. We discuss the vulnerabilities of
these knowledge components that can be exploited by different privacy attacks and
discuss possible defensive strategies to mitigate such attacks in 
Section~\ref{sec:att_dsl}.

\subsection{Distributed Recurrent Neural Networks}
Recurrent Neural Networks (RNNs) are a type of supervised machine learning technique
that conducts the training process using a linear time-invariant system approach ~\cite{zaremba2014recurrent}.
An RNN's distinct architecture enables its AI systems to retain information from previously seen data or calculated parameters, simplifying the process of training with many parameters~\cite{zaremba2014recurrent}.

\textcolor{black}{The typical structure of an RNN model includes an input layer, a hidden layer, an output layer, and connections among neurons. The layers of the RNN are organized into \emph{time steps}, where each time step represents a point in the input sequence. The connections between neurons allow information to be passed from one-time step to the next. The connections between the nodes in the input and hidden layers are parameterized by a weight matrix. The weights in the hidden layer represent recurrent connections within the hidden layer across different time steps.}

Training an RNN model consists of two methods, (1) forward propagation and
(2) backpropagation. Forward propagation, also known as the forward pass, is the process of calculating and storing the intermediate variables, including the output, for a neural network in a sequential manner, starting from the input layer and moving toward the output layer~\cite{catena2022distributed}. Backpropagation is an algorithm that is used to train neural networks by calculating the gradient of the network's parameters using gradient descent. It involves propagating the error back through the network in order to update the weights and biases in a way that minimizes the overall error. This method traverses the network in reverse order, from the output to the input layer, according to the chain rule from calculus, or uses gradients from timestamp T to be 
propagated all the way back to timestamp 1.

There are two main distributed RNN architectures that are used for distributed learning: distributed Long Short-Term Memory (LSTM) and distributed Gated Recurrent Units (GRU). These architectures involve maintaining and training a neural network locally for each Virtual Network Function Instance (VNFI). During each training iteration, the weights of the fully connected layers are exchanged among the neural networks using a consensus strategy. This allows the distributed RNN to learn and improve its performance over time~\cite{catena2022distributed}. 

Generally, the distributed LSTM mechanism is used in applications that involve language
modeling and machine translations~\cite{dong2022short}. However, due to the need for high efficiency,
the general LSTM mechanisms can be extended to the distributed paradigm. There are several architectures that enable the use of distributed LSTM networks, but one of the most prominent is the distributed computation of mini-batches. This involves dividing the training data into small batches and distributing the computation of each batch across multiple nodes or devices.\par
One example of this type of distributed LSTM training is \emph{Sequence Discriminative Distributed Training of LSTM}~\cite{sak2014sequence}. This method aims to improve the performance of distributed LSTM networks by optimizing the sequence-level objectives of the model.
This mechanism uses an asynchronous stochastic gradient descent mechanism to facilitate distributed learning. The process of this mechanism can be summarized as follows. Local workers partition an utterance (speech) and use their local data to analyze it one at a time. For each utterance, the speech module retrieves the necessary model parameters from the parameter server and uses them to calculate the probability of each state for each frame of the utterance. The module also decodes the speech to generate lattices and gather statistics on occupied states. The training task then continues with cross-entropy training until sufficient statistics are obtained~\cite{sak2014sequence}.

The distributed LSTM architecture involves the use of virtual network function instances (VNFIs) to connect with remote participants. Each VNFI has an agent that trains an LSTM-based neural network using monitoring data collected in VNFI. The architecture proposed by VNFI includes the use of agents and VNFIs to train the neural network ~\cite{park2020machine}. In ~\cite{park2020machine}, each agent or worker must share the weights of its neural network with the other agents through a regulated process in order to reach a global optimum. The majority of the knowledge components in this architecture are the weights of the neural network, which can take various forms. These weights are derived from different parts of the model participants and contribute to the final logic of the model.
In some cases, resource allocation can also be a significant factor in the efficiency of the model, and additional knowledge components may be shared to optimize resource allocation ~\cite{sak2014sequence}.

Distributed Gated Recurrent Units (DGRUs) are a form of distributed artificial neural network architecture utilized in natural language processing and computer vision applications. GRUs are a type of variation RNNs and handle the processing of sequential data. The Gated Recurrent Unit (GRU) architecture involves the use of two sets of weights that are updated during the learning phase to control the current state and activation based on the current input and previous memory. These weights are important in determining the behavior of the model~\cite{cho2014properties}. When this mechanism is extended to a distributed setting, it is known as DGRU. In a DGRU network, multiple GRUs are connected and operate in parallel, allowing for faster and more efficient processing. In the current literature, this is achieved by combining GRU and FL strategies. GFCL ~\cite{talpur2022gfcl} is a comprehensive example to demonstrate the integration of GRUs with FL to enhance the performance of detecting data poisoning attacks. This GFCL architecture can predict the characteristics of future data samples by analyzing a current data sample in a sequence-to-sequence regression manner. And GFCL uses FL architecture to facilitate communication between the end devices and the network. Another approach that uses above discussed distributed learning techniques is Privacy-Aware Human modeling ~\cite{ezequiel2022federated}. Privacy-Aware Human modeling focuses on creating human behavior models while preserving personal privacy. It is accomplished by building models without sensitive personal data, using anonymized or de-identified data, and implementing security measures. Another approach within this strategy involves using spatiotemporal event data for modeling, providing insight into human behavior without collecting personal data.

 \subsubsection{The Knowledge Components} \hfill\\
 Distributed Recurrent Neural Networks consist of various knowledge components, including those from Long Short-Term Memory (LSTM) and Gated Recurrent Unit (GRU) architectures. These architectures have different ways of sharing knowledge. To comprehend the knowledge components of distributed RNNs, it is necessary to examine LSTM and GRUs individually.

In a distributed LSTM network, there are several memory blocks called cells that transfer information between each other through  \emph{cell states} and \emph{hidden cell states}. These cell states are important components that facilitate the LSTM architecture. Additionally, there are two main memory components in an LSTM structure that contribute to the model:  \emph{memory from the previous cell} and  \emph{memory of the current cell}. These are viewed as the core knowledge components of distributed Recurrent Neural Networks (RNNs) and must be shared for distributed learning to take place. In more dependable architectures that prioritize reliability over efficiency, it has been noted that the \emph{active state of the current cell} and certain weights related to the forget gate should be shared among participants as well. These extra components are considered to be knowledge components in distributed RNNs. ~\cite{lyu2019road}.

The  \emph{knowledge component} in the distributed GRUs~\cite{talpur2022gfcl} is slightly different from the distributed training for LSTM because the  \emph{knowledge} can be solely paired with the  \emph{logits} and  \emph{gradients}. We observed that the GRU node consists of three main components: input data acquisition and pre-processing, and these local nodes share  \emph{model updates} with a global node to make predictions. So when we look at the knowledge-sharing phase, we can identify these \emph{model updates} are the  \emph{logits} of each local node. Furthermore, there will be  \emph{gradients} shared as knowledge components to facilitate federated learning. The sharing of knowledge makes distributed GRUs useful in cybersecurity as they introduce less complexity due to their fewer gates.
 
\subsection{Attacks on Distributed Supervised Learning}
\label{sec:att_dsl}
Upon investigating attacks on the knowledge in a Multilayer Perceptron (MLP), we have determined that the main knowledge components of a distributed MLP are \emph{gradients}, \emph{parameter distribution}, and \emph{aggregators}. We have identified vulnerabilities in these knowledge components and various attacks. Although attacks against distributed MLPs are still in the early stages, we propose a threat model that incorporates diverse vulnerabilities to exploit a distributed MLP model. This threat model considers the general MLP architecture and its related knowledge components.\par
In a distributed learning setup, the activation functions of the model and some hidden layers are split among the model participants. As a result, the gradients, parameter distribution, and aggregation information will be situated in the layers. If we consider the global model to be honest but inquisitive, it may be susceptible to model memorization attacks, as proposed by ~\cite{10.1145/3133956.3134077}. Using attacks like  \emph{model memorization}, the attackers will be able to conduct attacks in both black-box and white-box scenarios and will be able to steal  \emph{knowledge components} and then replace them with malicious content. Song et al. ~\cite{10.1145/3133956.3134077} have demonstrated that this attack will be more impactful than any other attack on MLP.
The knowledge components in distributed CNNs were mainly identified as the layers due to their widely used layer pipelining architecture. However, as we discussed in section \ref{sec:DCCNN}, the  \emph{skewness factor},  \emph{demand layer}, and  \emph{partitioning point} are also some sub-knowledge factors that support the State-Of-The-Art (SOTA) smart layer placement mechanisms that can be seen in the current industrial applications.\par

To create a more thorough understanding of the defensive strategy for knowledge sharing in distributed CNN, we have developed a threat model that takes into account all of the various  \emph{knowledge components} and prioritizes the  \emph{layers} based on the types of attacks currently being carried out on existing distributed CNN implementations. Before proposing the threat model, we  first, explore the vulnerabilities associated with these components and then demonstrate that our threat model that exploits these vulnerabilities. \par
When looking into the knowledge that consists of the layers, it can be easily intercepted by attacks like   \emph{model inversion  timing attacks}~\cite{chen2022inversion}, which is a combination of inversion and timing attacks in order to reconstruct the training data used to create a machine learning model. This is done by making repeated queries to the model and using the response time to infer what the training data may have looked like. The timing aspect of this attack involves using the time it takes for the system to respond to the queries to gather information about the system or data being processed. Chen et al. ~\cite{chen2022inversion} have successfully implemented an attack using the timing attack in a novel aspect by combining side-channel attacks in the Security Operations platform for the model inversion. The knowledge components, such as the  \emph{skewness factor} and the  \emph{layer partitioning point}, are variables that can be easily intercepted in the backpropagation phase where knowledge sharing happens. Data skewing is one of many reasons that lead to data poisoning attacks ~\cite{miao2018towards}. Miao et al.~\cite{miao2018towards} demonstrated the impact of data skewing toward model poisoning attacks. \textcolor{black}{Data skewing refers to the phenomenon of having imbalanced data distribution in a  data set. This can bring a variety of vulnerabilities into the learning model, such as bias towards the more weighted class, where underrepresented data classes would neglect. This scenario will eventually lead the model to overfit, where the model becomes too specialized to the characteristics of the skewed data.} Furthermore, Miao et al. ~\cite{miao2018towards} present how the  attackers aim to manipulate  \emph{skewness} values to certain target values and then against a crowd-sensing system empowered with the truth discovery mechanism. When considering the  \emph{partitioning point} knowledge component, by obtaining the feature data partitioning points, the attackers are capable of initiating membership inference attacks ~\cite{shokri2017membership}.\par 
By integrating these SOTA scenarios, we generated the following threat model. In our threat model, we consider an honest but curious global model. We also consider a typical distributed CNN network that establishes its distribution through backpropagation for model updates and forward propagation for aggregation\cite{lindell2020secure}. To cover both white-box ad black-box scenarios, we consider an instance where a feature estimation attack is being initiated. By implementing a feature estimation attack, the adversary might be able to obtain certain features like the  \emph{skewness factor} and  \emph{layer partitioning point}, leading the model to be compromised. We also consider the adversary can initiate a model inversion attack. In that case, the adversary will be  able to  adjust the weights and obtain the features for all classes in the network, unraveling the hidden layers.\par
Based on the vulnerabilities in various distributed learning settings we have discussed above, the knowledge components related to the "neural network update information" category are the most susceptible to different types of attacks. As a result, this category can be identified as the most vulnerable knowledge category.

\subsection{Defences on DSL}
\label{sec:Def_DSL}
Effective defense strategies against the threat model mentioned are scarce. Hence, it's crucial to factor in defensive strategies when exploring ways to reduce this type of attack.\par 
Examining the weaknesses that can undermine the integrity of a distributed MLP model, it becomes evident that model memorization and communication of \emph{knowledge components} between participants can render the model vulnerable. To address these vulnerabilities, two defense strategies are proposed: securing the communication channel and securing the shared knowledge components. Our emphasis will be mainly on securing the knowledge components as the communication medium between ML models can be complex and dynamic, making it challenging to anticipate and defend against all potential security risks, thereby making it hard to guarantee privacy and security fully. As previously discussed in section \ref{sec:dsl_mlp}, the main knowledge components in a distributed MLP are \emph{gradients},  \emph{parameter distribution}, and  \emph{aggregating parameters}.\par
Several defensive mechanisms can be used to preserve the privacy of these knowledge components. . Those methods will mainly consist of techniques based on encryption and aggregation with differential privacy. Aono et al.~\cite{aono2017privacy} proposed a secure framework that employs additive homomorphic encryption for storing encrypted gradients on a cloud server. This encryption method allows for computation across the gradients. While its primary purpose is to secure gradients, it can also be utilized for modeling parameter distributions and aggregators.
Despite the implementation of differential privacy and gradient compression for label protection, our proposed threat model suggests that an adversary can still make accurate predictions.~\cite {liu2022clustering}. When specifically looking into the defensive strategies that can be implemented on  \emph{layers},  \emph{skewness factor}, and  \emph{partitioning points}, the most common defensive strategy is obfuscation, but it introduces a considerable negative impact on the model's performance and interpretability~\cite{geng2020tight}. This is a critical issue that should be addressed in future research. When considering the preservation of privacy for knowledge components classified as \emph{weights}, there are five different defensive strategies that can be incorporated into a distributed supervised learning scenario, namely, differential privacy ~\cite{xie2021differential}, homomorphic encryption ~\cite{wibawa2022homomorphic}, FL ~\cite{asad2021federated}, secure muti-party computation~\cite{atallah2001secure} and the use of trusted execution environments~\cite{narra2019privacy}.

\section{KNOWLEDGE SHARING IN DISTRIBUTED UNSUPERVISED LEARNING}
\label{sec:knowledge_sharing_in_udl}
Unsupervised learning is a technique for finding patterns in unlabeled data without human direction. Advanced unsupervised algorithms can uncover hidden patterns in massive datasets, improving accuracy and reliability for classification and prediction tasks. These algorithms examine  unlabeled input data, which has not been organized into specific categories. Instead, they don't have a pre-defined output and focus on discovering relationships and patterns within the input data. The training of the machine learning model utilizes unlabeled input data. Initially, it interprets the raw data to uncover hidden patterns. Subsequently, k-means clustering and other suitable algorithms are applied to group the data objects based on their similarities and dissimilarities.~\cite{liu2022clustering}.\par 

There are two main categories of algorithms that can be used in unsupervised machine learning: Clustering and association. Clustering algorithms are used to identify inherent groupings within a dataset, while association algorithms are used to discover specific rules that define the behavior of a dataset. The choice of algorithm to use depends on the application's specific needs. Moreover, unsupervised algorithms can be used for Dimensionality reduction tasks seen in Principal Component Analysis (PCA) and autoencoders.  The architecture of unsupervised learning is determined by selecting and implementing the appropriate algorithm, which groups the data objects based on their similarities and differences.  When unsupervised machine learning algorithms are used in a distributed environment, it is referred to as a  \emph{distributed unsupervised learning} (DUSL) scenario. DUSL can improve these unsupervised algorithms' effectiveness, accuracy, and scalability, particularly when working with large datasets. This  \emph{DUSL architecture} can be categorized into  three main categories, namely distributed generative adversarial networks,  \emph{distributed autoencoders}, and  \emph{distributed SOMs}. Based on the working schemes of these categories, the  \emph{knowledge sharing} will be unique, so the vulnerabilities associated with them as the defense mechanisms.
\begin{figure}[t!]
    \centering 
    \includegraphics[width=0.5\linewidth]{ {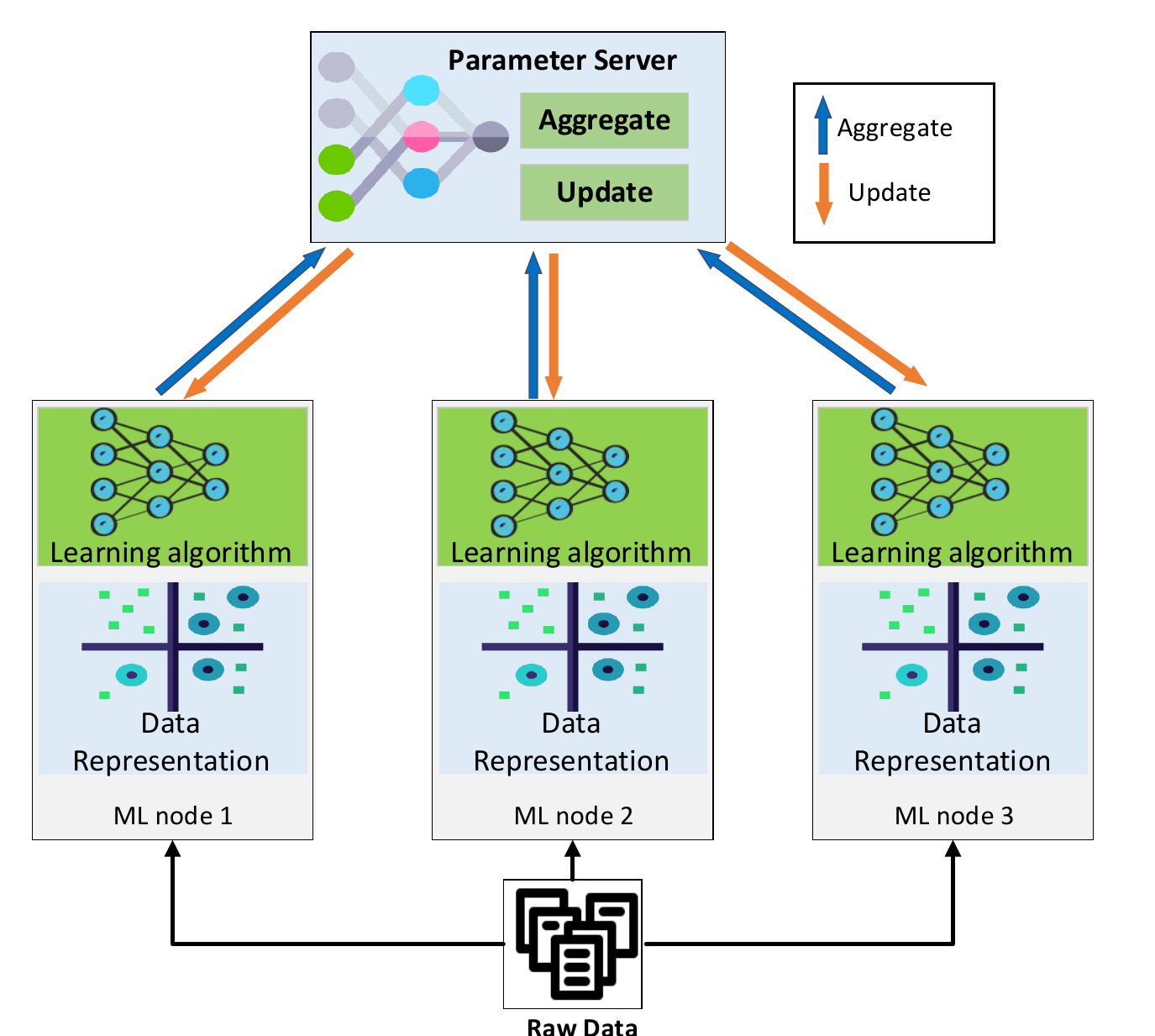} }
    \caption{Knowledge sharing in unsupervised learning.} 
	\label{Fig: Knowledge sharing in DUSL}
	\vspace{-1mm}
\end{figure} 
\subsection{Distributed Generative Adversarial Networks}
A Generative Adversarial Network (GAN) is a two-part machine learning model structured like a two-player game. The objective is to minimize the loss function, which compares generated data to actual data ~\cite{creswell2018generative}. GANs usually consist of a generative model and a discriminator model. The generative model is trained to replicate the distribution of the training data. In contrast, the discriminator model determines whether a sample is generated by the generator or taken from the training data. The generator is trained to maximize the discriminator's false positive rate. The above discussed is the typical training procedure of local GAN networks. When looking into the  \emph{distributed GANs}, once the training of the distributed GAN is brought forward to the edge of the network. Similar to other distributed learning architectures, the learning for the GAN can be achieved via data parallelism and model parallelism architectures.\par
As the focus is on \emph{knowledge sharing}, this discussion will only cover cases where model parallelism or a combination of data and model parallelism (hybrid parallelism) are implemented. In GANs, the parallelism between end nodes can be achieved through two main approaches. The first is to employ multiple generators across the network to produce synthetic data and a single discriminator located in the global model to assess the similarity between the synthetic data and actual data labels. The second approach involves integrating the Private Aggregation of Teacher Ensembles (PATE) framework with GANs. This mechanism is introduced as PATE-GAN ~\cite{jordon2018pate}. Although this PATE-GAN ~\cite{jordon2018pate} is introduced as a privacy preservation mechanism due to the way it achieves synthetic data generation using  several teacher discriminators and one student discriminator. Using this mechanism, the authors of PATE-GAN developed a novel approach to generate  differential private synthetic data with strict, deferentially private guarantees. Additionally, some applications utilize multi-generator models to attain distributed learning. \par 
MG-GAN ~\cite{dendorfer2021mg} is a multi-generator GAN framework for pedestrian trajectory prediction. It effectively reduces the number of out-of-distribution samples compared to using a single-generator approach by using specialized distributed generators for sampling. Each generator is trained to produce synthetic data following a specific trajectory distribution towards a primary mode in the scene. Additionally, a second network is trained to categorize the generator distribution based on input data regarding scene dynamics.                       

\subsubsection{The Knowledge Components} \hfill\\
\label{sec:The Knowledge Components}
Since distributed GANs should achieve their inference in collaboration with generator and discriminator nodes, its knowledge-sharing phase is somewhat complex and comprehensive. Once the specific knowledge components are shared during the communication phase, it will be straightforward to intercept them. When looking into these knowledge components, it would be better to identify the distribution of the generator and discriminator nodes. Upon examination of the knowledge explicitly shared in the distribution of generator nodes, it was found that the majority of the model's knowledge is related to instances where outputs multi-generators collaborated. When looking into distributed discriminator-based  GANs, Hardy et al.~\cite{hardy2019md} discuss distributed GANs over a set of worker nodes. To overcome the computational complexity of workers, they have a single generator in the system hosted by the parameter server, which is made possible by a peer-to-peer communication pattern between the discriminators to the workers. The authors present the knowledge as the   \emph{Learning iterations} that comprise a multiplication of  \emph{batch size} with the  \emph{object size} of the data sample,  \emph{Error feedback}, which can be identified as a multiplication of batch size and the object size, Message size.\par
The approach in ~\cite{gao2019hyperspectral}  involves using multi-Discriminator GANs with distributed discriminators for hyperspectral image classification. It optimizes the GAN structure through ensemble learning. The process starts with preprocessing the hyperspectral image to obtain spatial-spectral samples, which are then used to train the generator. The authors use majority voting to determine if the generated samples are real or fake, and the voting scores from multiple discriminators guide the sample generation. The knowledge-sharing phase of ~\cite{gao2019hyperspectral} is completely different from the ~\cite{hardy2019md} because the knowledge components described in this instance are discriminating probability of a category. This mechanism~\cite{hardy2019md} uses the results of arithmetic, geomatic and harmonic averaging and then uses a SoftMax as the classifier.\par 
When presenting muti-generator based distributed GANs, MG-GAN~\cite{dendorfer2021mg} demonstrates similar architecture to distributed  discriminator approach presented by Hardy et al.~\cite{hardy2019md}, but the main difference is instead of using distributed discriminators, this approach uses a set of distributed generators. The  MG-GAN ~\cite{dendorfer2021mg} approach uses multiple generators that are specialized in different modes and are able to learn to generate samples based on the scene observation utilizing these generators. Upon the investigation of the knowledge components of this architecture, it can be identified that this MG-GAN ~\cite{dendorfer2021mg} does not share weights as a knowledge component. Instead, it consists of an LSTM decoder that is initialized using the classification features and a randomly generated noise vector. So we can identify that the distribution will be established mainly using the knowledge  \emph{logits} that the LSTM decoder will generate. Multi-level Generative Adversarial Network (M-GAN)~\cite{hoang2018mgan} is a type of generative model that uses a hierarchical structure of multiple generators and discriminators to generate images at different scales. M-GAN ~\cite{hoang2018mgan} examines the use of multiple generators in a GAN configuration to address the model collapse issue. This is achieved by using the generators in this system to produce samples that are intended to match the training data distribution. The discriminator is responsible for determining whether a sample is actual data or generated by the generators, and the classifier specifies which generator a sample was produced by. The knowledge components in this technique ~\cite{hoang2018mgan} are the  \emph{parameter distributions} except for the input layer. Most other research on distributed GANs via distributed generators is based on or facilitates deep reinforcement learning.\par
\begin{figure}[t!]
    \centering 
    \includegraphics[width=0.65\linewidth]{ {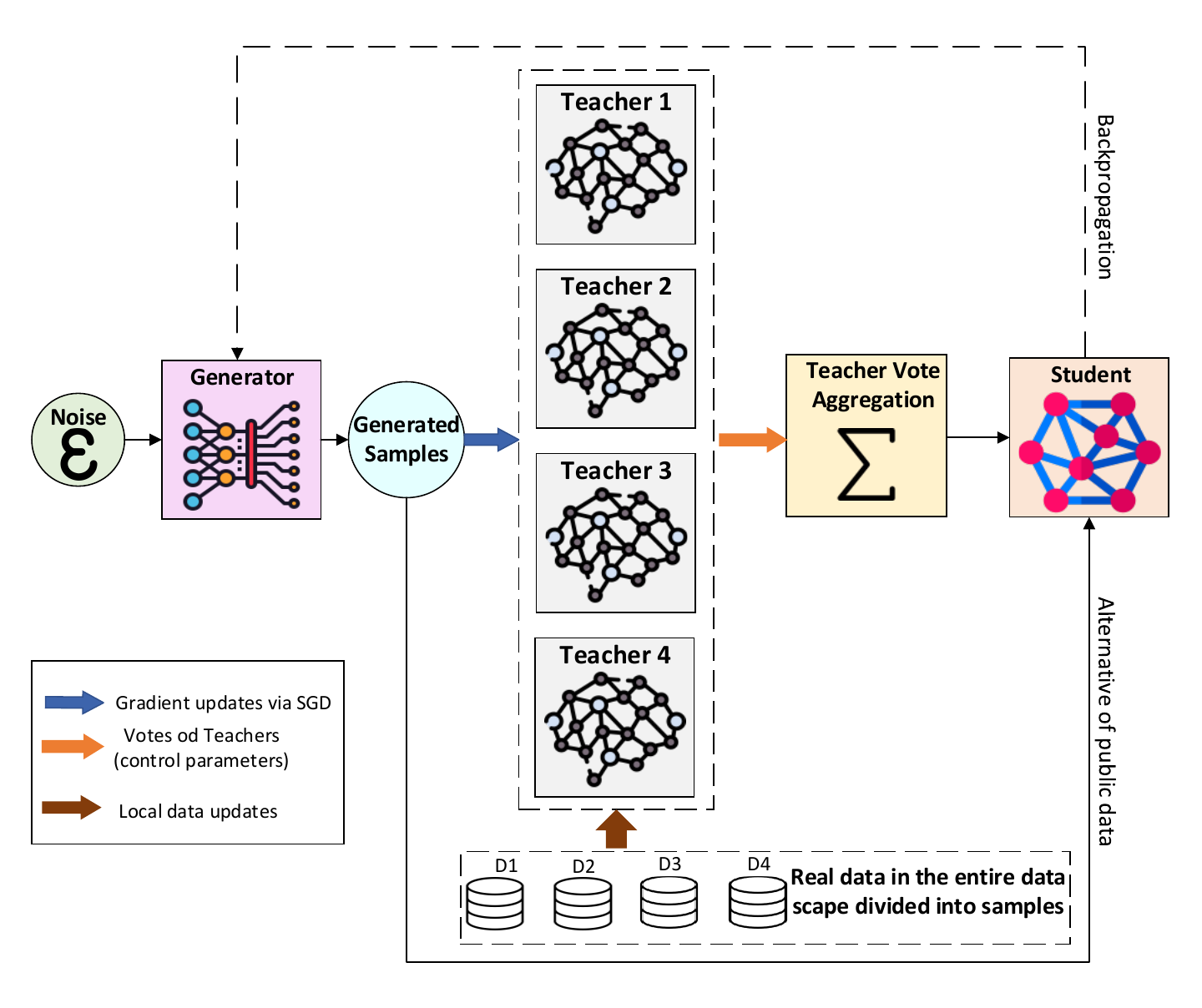} }
    \caption{The Learning architecture of distributed learning through distributed discriminators.} 
	\label{Fig: Knowledge sharing in PATE-GAN}
	\vspace{-1mm}
\end{figure}

Shi et al. ~\cite{shi2021distributed} propose a new approach that leverages the concept of reward shaping in reinforcement learning (RL). They suggest replacing the Earth-Mover distance with the Inception Score as the reward value for the controller in a single-level architecture search method. Each cell model is assigned a reward value. So the knowledge components that can be identified in this work are the  \emph{Controller parameter},  \emph{Policy gradient}, and the  \emph{weights} of the shared generator. When looking into another article that describes distributed GANs achieved via distributed generators, Zhang et al. ~\cite{zhang2021distributed} present a novel framework that is proposed to enable air-to-ground channel modeling over millimeter wave (mmWave) frequencies in an uncrewed aerial vehicle (UAV) wireless network. Their approach is to  collect mmWave
channel information allowing each UAV to train a local channel model via a GAN. The knowledge-sharing phase is mainly facilitated by generated wave samples which are the logits of the distributed generators and gradients that will be handled during the backpropagation. When summarizing the knowledge aspect of the distributed GANs, we identified that the knowledge components would be different from scenario to scenario and the distributed GAN implementation, as shown in Figure \ref{Fig: Knowledge sharing in PATE-GAN}. So it was identified that the attacker would have a broad attack surface to attack the distributed GANs.\par
When discussing the knowledge components of PATE-GAN ~\cite{jordon2018pate}, we must consider the three main components and their specific roles. Generally, the role of the generator remains similar to general GAN, so it does not produce any  \emph{knowledge}. It just  generates synthetic data by adding noise to the actual data. The  \emph{knowledge components} come to the surface when this generator updates the teacher components using Stochastic Gradient Decent (SGD), where  \emph{gradients} can be identified as the knowledge. The process of the distributed discriminator utilizing the PATE mechanism involves two components: teacher and student discriminators. The teacher discriminators primarily classify the synthetic data produced by the generator and make decisions based on their classification results. The teacher entities will determine the distance between the real data and the synthetic data generated by the generator. These distances will be sent to the  \emph{student discriminator} as  \emph{votes}, which we identify as knowledge components of  \emph{control parameters} since they are responsible for determining the right balance between losses of the synthetic data coming through the teacher discriminator and the actual data. 

\subsection{Distributed Autoencoders}
Autoencoders are mainly used in domains ranging from feature selection to information retrieval~\cite{han2018autoencoder}. In the current research landscape, the main use of autoencoders is for dimensional reduction tasks in industry applications ~\cite{han2018autoencoder}. The general autoencoder architecture has three main components: an input layer, a hidden layer, and an output layer. The encoding process involves using the input variables to generate a representation in the hidden layer. Then the model will succeed once the decoding starts projecting the hidden layer to the output layer. So typically, an encoder will construct the input data into a representation and then will use this representation to obtain the output. Although this process is successful in small-scale networks, this process might be inapplicable in large-scale processes where multiple operating units possess their own unique features while being interconnected with other units. Because, unlike the unique features that affect the local units, when global features that interconnect model participants are altered, the entire model can be compromised. Independent features primarily affect the local units, with a minimal global impact. The entire process can be faulty when the global features that connect various operating units are altered ~\cite{li2021distributed}. Distributed autoencoders have emerged as a solution to this problem. They have been used in a wide range of applications to transmit data without compromising privacy ~\cite{zhang2019privacy}.\par
When looking into these distributed autoencoders, it can be identified that the  \emph{encoder} is the main component that will be distributed through the network where the decoder will be positioned in the global model. Further, there are applications where distributed recurrent autoencoders are used; DRASTIC ~\cite{diao2020drasic} is such an application where a distributed set of encoders will be trained on one collaborative decoder and correlated data sources. These distributed encoders can perform as well as a single encoder trained with all data sources together. There are applications where both encoders and decoders are distributed, but it is worth mentioning that in those instances, utility is the highest priority over privacy. Lu et al.~\cite{lu2022distributed} have used distributed encoder/decoder model for spatial pattern recognition. When looking into their working scheme to encode, it mainly involves converting a message into a cell seeding configuration and then allowing the cells to grow and form a colony pattern.

\subsubsection{The Knowledge Components}\hfill\\
In distributed autoencoders, the knowledge being shared as it is typically the representation of the input data or the representation of the hidden layer. In these scenarios, the knowledge will be in data representation snippets. To represent a collective data set, most of the models seek to identify a pattern that captures all of the data within the scope since these patterns are transferred from one model to another to attain collaboration.\par
When looking into the knowledge components transmitted in a semi-supervised autoencoder setting, we identified that most of the knowledge components were similar to general d autoencoders, which are  \emph{logits}, knowledge components origin containing the information of encoding and decoding results. But due to the  distributed architecture, some foreign knowledge components come into play. These other components introduced by the classifiers are responsible for integrating supervised learning into the unsupervised encoder paradigm. Due to this integration, the knowledge will be the  \emph{latent distribution's mean} and the  \emph{latent distribution's variance} that will be used for scaling purposes.

\subsection{Distributed SOM}
Self-Organizing Maps (SOM) are neural networks that use unsupervised competitive learning to improve their accuracy. In this process, nodes in a network compete for the opportunity to process certain inputs, leading to a specialization of the nodes and reduced redundancy. As the network iterates, it becomes more closely aligned with the training data. However, typical SOMs are used for visualization and exploratory data analysis of high-dimensional datasets. SOMs also group similar data points, providing a way to visualize clusters in the data. This feature allows SOMs to reduce the dimensionality of the data and highlight similarities among data points. When applied to a distributed network, SOMs can be used to address issues related to performance bottlenecks in distributed models. SOMs achieve their working scheme via the Kohonen algorithm~\cite{kohonen2007kohonen}. When this Kohonen algorithm is modified into a parallelization, distributed SOMs can be achieved. The goal of Distributed SOMs is  to overcome the current drawbacks of the data platforms.\par
Most of the SOM approaches are based on  clustering the data at data sources~\cite{phan2017distributed}. Distributed SOMs achieve this by moving computations to the data sources. For this, the SOM algorithms are decomposed to perform locally on the nodes without transferring data in the network~\cite{phan2017distributed}. This helps to reduce the execution time and the network traffic. This is achieved by building upon earlier general techniques for running data mining algorithms on distributed data sources in parallel. In the current research, there are several developments in the approach mentioned above. DSOM ~\cite{phan2017distributed} is a framework that allows application programs to access objects across address spaces. That is, application programs can access objects in other processes, even on different machines. The location and implementation of an object are hidden from a client, and the client accesses the object (by method calls) in the same manner regardless of its location. \par
The current developments in distributed SOM are not just limited to the parallelization of the algorithm. Some works integrate FL to facilitate distributed SOM ~\cite{kholod2021parallelization}. By integration of federated learning, the authors of ~\cite{kholod2021parallelization} enable the execution of parallel algorithms on the distributed data sources, taking into account the kind of data distribution on the nodes. This method facilitates distributed learning via the SOM clustering algorithm, which transforms SOM into a parallel implementation that performs major calculations at the nodes rather than transferring data to a central node. Because of the aggregation into a single point, the SOM can be the bottleneck of detection throughput or even a target of DoS attacks. Using this hypothesis, the authors of ~\cite{kim2015distributed} have presented a mechanism to distribute single SOM to multiple points. The multiple SOMs are separately running at the corresponding points, and they have come up with a Distributed SOM as a solution to make the existing SOM-based approach more scalable and robust against DoS attacks demonstrating the broad usability of the Distributed SOM.

\subsubsection{The Knowledge Components} \hfill\\
In the above, we discussed the distributed SOMs via three distinguishable implementations. Now, we are looking into the knowledge components that were associated with them. We identified that they all are based on one architecture but differ based on their use. So to identify the  \emph{knowledge} components more closely, it is essential to go through the process of DSOM. Generally, the DSOM process consists of three major steps, \emph{Initializing}, where each DSOM is trained with initial  input data, then \emph{using the initial map}, each DSOM operates separately on receiving its own input data, it reorganizes its map and classifies the input data. And by \emph{obtaining the weighted sum of SOMs}, if each DSOM continues to operate separately, the map deviation becomes so big that each DSOM can make a different result even for the same input. \newline
In the DSOM process, multiple separately trained maps are combined to create a  \emph{merged SOM}, which serves as the global model in the distributed network. We have found that the  \emph{knowledge} in a distributed SOM mainly consists of merged classification logits.

\subsection{Attacks and Vulnerabilities on DUSL}
In the previous sections, we presented the main unsupervised learning architectures that have been mainly used in modern applications. With a thorough outlook on their initial architecture, we can identify that knowledge sharing happens in different forms throughout all the architectures, although their initial architecture is the same. The vulnerabilities of the knowledge components will be identified, assuming this knowledge sharing can be intercepted at the communication phase. Our assumption is supported by the SOTA attacks against these generally distributed machine learning architectures.\par
The main  \emph{knowledge} components that can be identified in DUSL are the  \emph{logits},  \emph{weights},  \emph{controller parameters}, and  \emph{gradients}. Unsupervised learning does not rely on  \emph{logits},  \emph{controller parameters}, or  \emph{weights} to make predictions or classify data. It simply aims to find patterns in the data without using labeled examples. Therefore, unsupervised learning is not vulnerable to attacks applicable to DSL architectures. However, if the  \emph{policy gradients} component of the model is trained on malicious data, it may become biased in favor of the malicious party, leading to inaccurate results. Nevertheless, the secondary knowledge components, which are optimized for more accurate predictions, can be intercepted easily at the communication phase since such knowledge components are transferred as basic information. Therefore, the secondary knowledge components such as  \emph{batch sizes},  \emph{object size}, distribution means, and variances introduce vulnerabilities by allowing attackers to gain access to sensitive data or model behavior. Intercepting information during the communication phase can allow an attacker to interfere with or alter the machine learning process. These vulnerabilities can eventually lead to  attacks such as  \emph{Feature estimation},  \emph{model reconstruction}, and  \emph{model poisoning attacks}.\par
When examining the knowledge components that are susceptible to attacks in distributed unsupervised learning, we have found that all components related to each category discussed in section \ref{sec:knowledge_sharing_in_dl} is equally impacted by the complexity arising from the coordination and communication required between multiple machines or nodes in the network to maintain consistency and reliability in the learning process.

\subsection{Defences on DUSL}
When looking into the defenses that can be implemented in distributed unsupervised learning, there is not much research that discusses defending against attacks. Since we are mainly concentrating on the  \emph{knowledge sharing} and vulnerabilities associated with them, we were able to specify the defensive techniques that can be used on the  \emph{knowledge} components to ensure their privacy of them. Considering the model's utility is essential when examining defensive strategies for knowledge components. Some defensive strategies are only applied in distributed ML settings to maintain maximum utility while protecting knowledge components. Distributed GANs have revealed various knowledge components. Those \emph{knowledge components} can be identified as the  \emph{logits}, \emph{batch sizes}, \emph{error feedback}, \emph{controller parameters}, and \emph{gradients}. Furthermore, in SOMs also, the main knowledge was \emph{logits}, while distributed autoencoders facilitate their distributed training via latent state distributions of  \emph{variance parameters} and  \emph{mean parameters}. In order to defend these knowledge components, we can suggest the privacy preservation techniques like  \emph{sharing fewer gradients} ~\cite{phong2017privacy}, where only essential gradients would be shared at the updating phase, incorporating differential privacy ~\cite{geyer2017differentially}, and where some noise will be added to the gradients or parameters. Further, we also identify \emph {Dimensionality reduction} ~\cite{shokri2017membership} as a defensive strategy where participants only rely on predefined parameter sets.

\section{KNOWLEDGE SHARING IN DISTRIBUTED SEMI-SUPERVISED LEARNING}
\label{sec:knowledge_dssl}
Semi-supervised learning addresses the challenge of classifying large, heterogeneous, and difficult-to-label data commonly found in modern edge and cloud device services. This method is widely used in industry to process vast amounts of data obtained from smart devices like IoT sensors, which are abundant but hard to classify due to their diversity. In current applications, semi-supervised learning is a key approach for handling such data ~\cite{zhu2009introduction}. This semi-supervised learning architecture combines classification and clustering algorithms to group data based on similarity and then uses clustering algorithms to determine the relevance of the data samples. The data can then be labeled and used for machine learning. It is clear that semi-supervised learning can be used for both inductive and transductive learning tasks ~\cite{wang2014large}. Based on these approaches, we recognize three major architectures of semi-supervised learning: semi-supervised GANS, distributed transformers, and distributed contrastive learning mechanisms. Next, we discuss each of these architectures in the following Sections \ref{sec:DSSGAN}, \ref{sec:DT}, and \ref{sec:CONSTRIVE}.

\begin{figure}[t!]
    \centering 
    \includegraphics[width=0.75\linewidth]{ {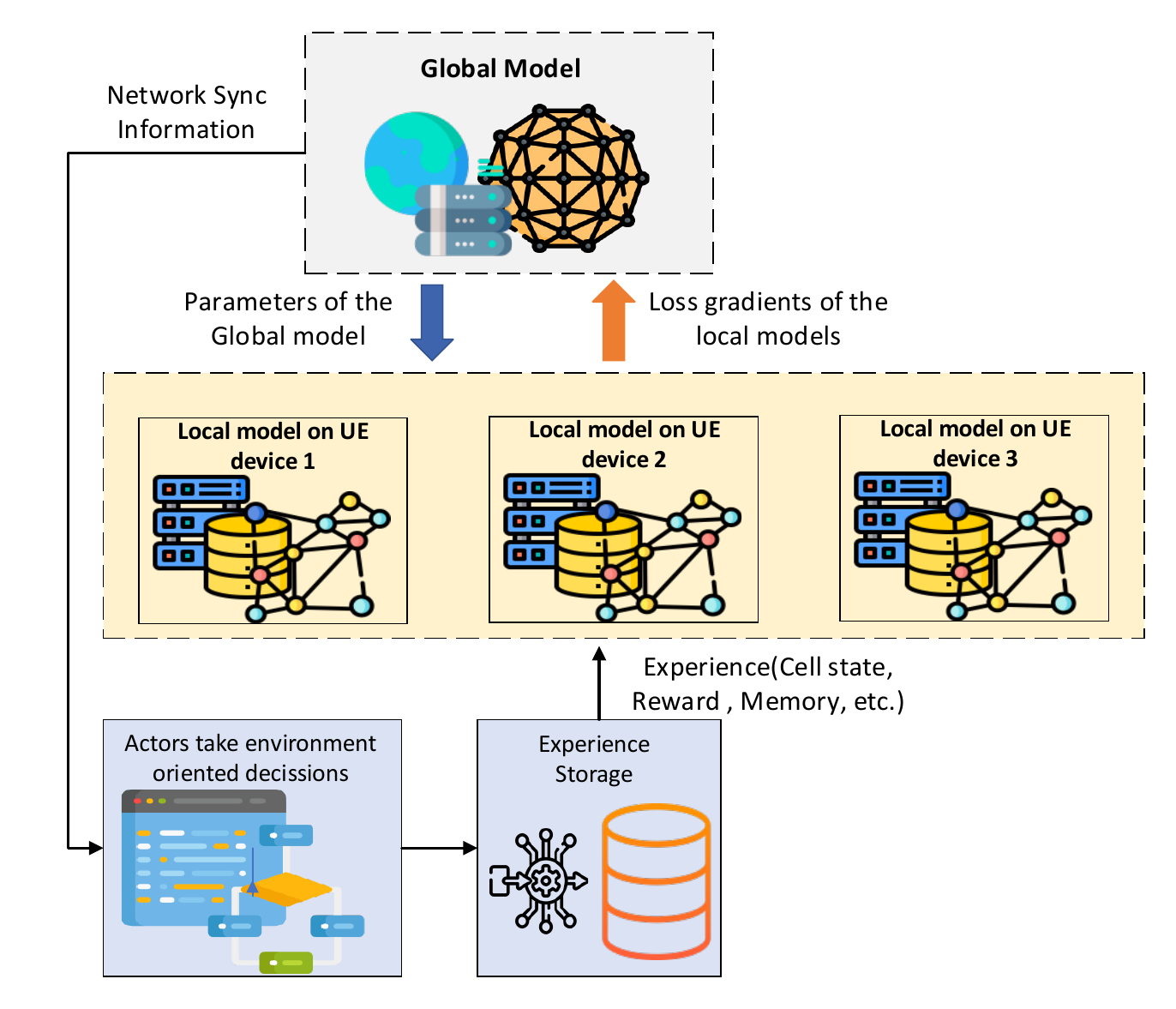} }
    \caption{Distributed Semi-Supervised Learning.} 
	\label{Fig: Knowledge sharing in DSSL}
	\vspace{-1mm}
\end{figure} 

\subsection{Distributed Semi-Supervised GANs}
\label{sec:DSSGAN}
It is important to note that traditional GANs are a type of unsupervised learning mechanism. In this section, we will examine modern implementations of GANs that integrate semi-supervised learning mechanisms to improve the performance of the model. And they can be implemented in a distributed environment, which leads to   \emph{distributed semi-supervised GANs(DSSGAN)}.In DSSGAN, the generator is trained to create data that resembles the training data, while the discriminator is trained to differentiate between real and fake data. In a DSSGAN, the generator and discriminator are trained through a process of competition. The generator tries to create data that the discriminator cannot tell apart from real data, while the discriminator tries to accurately classify the data as real or fake. The goal of a DSSGAN is to learn a model of the data that can generate realistic samples. These networks are often used in situations where there is a small amount of labeled data but a large amount of unlabeled data ~\cite{zhang2021semi}.\par
The classification of imbalanced data has become a widely recognized concern. One solution to this issue is to create new data for the minority group to meet the needs of the majority group in light of this method. Zhou et al. ~\cite{zhou2018gan} introduced a method that generates data using the GAN model and then initiates classification. This architecture is presented as a semi-supervised GAN. Authors in ~\cite{he2017generative} have expanded this idea by utilizing GANs to fully use labeled and unlabeled data and create a classification map for a more robust classification. When this method is applied to a distributed environment, this will be identified as a  \emph{distributed semi-supervised GAN}. Since these generators, discriminators, and classifier work as separate entities, we consider this learning mechanism a distributed learning instance. Figure \ref{Fig: Knowledge sharing in DSSL} demonstrates this idea clearly.

\subsubsection{Knowledge Components in Semi-Supervised GANs}\hfill\\
Semi-Supervised GANs achieve semi-supervised learning by combining typical distributed un-supervised GANs and supervised classifiers. So when considering the knowledge components of semi-supervised GANS, we should consider the  \emph{knowledge} that is being shared from the supervised and unsupervised architectures. Although some mechanisms split single-view data into multiple views and facilitate distributed semi-supervised learning ~\cite{wang2008random}, all the models that are being discussed in this  paper select knowledge components from distributed parties. We have identified two main components that contribute to the knowledge of this architecture. These components are the knowledge generated from the GAN and the knowledge generated at the classifier. When looking into the  \emph{knowledge} shared by the GAN, typically, it will be the  \emph{tangents of the data manifold} estimated by the generator of the GAN ~\cite{zhou2018gan} ~\cite{kumar2017semi}. In some instances where the classifier is placed in the global model, the distributed parties will share  \emph{gradients} and  \emph{parameters} as the knowledge ~\cite{kirienko2021distributed}. And also, we have identified that there are instances where distributed classifiers that facilitate the learning process of a GAN are placed in the global model ~\cite{guan2020gan}. In those instances, these distributed classifiers will share the classification results that we identified as \emph{logits} in relation to the knowledge.

\subsection{Distributed Transformers}
\label{sec:DT}
Transformers are neural networks that specialize in sequence-to-sequence tasks. They are commonly employed in natural language processing (NLP) for translation tasks~\cite{park2022multi}. In summary, the functioning of this mechanism involves identifying patterns in the input sequence and using those patterns to identify patterns in the output sequence. When transformers are used in a distributed environment, it is known as \emph{Distributed Transformers}. This is a type of neural network architecture presented by Viswani et al. ~\cite{vaswani2017attention} in machine learning to handle tasks involving sequence-to-sequence architectures. This type of architecture is highly scalable and can effectively process large datasets, making the learning process more efficient. These models have been applied to various fields, such as natural language processing, machine translation, and language modeling.\par
Transformers differ from encoder-decoder architecture due to the inclusion of an additional technical component called \emph{Attention}, which is responsible for deciding the prioritized components in an input sequence. Vaswani et al.~\cite{vaswani2017attention} identified that the implementations that initiate image classifications mainly used this distributed learning architecture. This is because of the large data file host that can be found in image recognition tasks. FeSTA ~\cite{park2022multi}  framework uses split learning with vision transformers ~\cite{dosovitskiy2020image} to achieve distributed learning. In ~\cite{park2019accelerated}, the authors have developed a shared task called Agonistic ViT, which is a type of encoder-decoder architecture based on the transformer model. This model is trained to process sequential data, such as natural language. In their approach, the transformer model is distributed on the server side as the global model, and task-specific convolutional neural network heads and tails are placed on the client side. This allows the model to be trained on multiple tasks simultaneously, relying on the main knowledge illustrated through the FESTA framework ~\cite{park2022multi}, which concentrates on privacy preservation using permutation. Since this method facilitates the improvement of single tasks under distributed multi-task collaboration, it is widely used in the healthcare sector ~\cite{park2021federated}. Pipe-Transformer~\cite{he2021pipetransformer} uses a general elastic pipelining and elastic training acceleration framework that automatically reacts to frozen layers by dynamically transforming the scope of the pipelined model and the number of pipeline replicas.

\subsubsection{Knowledge Component in Distributed Transformers}\hfill\\
Like other distributed semi-supervised learning approaches, transformer architecture often employs a classifier to its general unsupervised architecture to achieve semi-supervised behavior and make robust predictions in the presence of anomalies in the dataset. In order to perform these tasks with minimal computational cost, the collaboration between distributed entities will be conducted in a split manner, with all parties effectively communicating during the training phase. The knowledge components of the transformer architectures we have discussed in previous chapters typically include task-specific head and tail parameters between clients, which are sent to the aggregator. When examining the general design of these implementations, we can see that these components play a central role~\cite{park2022multi}.\par
Some architectures that are concerned with the weights and gradients associated with the convolutional heads and tails include a permutation module. This module helps to protect the confidentiality of the data being transmitted between clients by obscuring the weights that are sent by the clients and sending permuted data components to an aggregator ~\cite{park2021federated}. This mechanism is unique since it follows pipelining procedures. Due to this architectural indifference, we know that in pipelining processes used to achieve parallelism, two methods are being used, which are data parallelism and model parallelism. Park et al. ~\cite{park2021federated} use both of these methods to facilitate the objectives. Our investigation has identified that the knowledge components of this model are primarily found in the model parallelism phase. In the context of \emph{knowledge sharing}, we have identified that the \emph{gradients} that define the partitions endpoint and the \emph{immediate output} of specific partitions or the last multi-headed \emph{attention layer} are the primary knowledge components used in this model. These components are essential for effective knowledge sharing in this model.

\subsection{Distributed Contrastive Learning} 
\label{sec:CONSTRIVE}
Contrastive learning is a Semi-supervised  machine learning paradigm that encourages the learning of a general data set without labels by teaching the model about similarities and differences in different augmentations of the data points. The goal of contrastive learning is to minimize the distance between two data points with different augmentations in order to optimize the model. This is achieved through the use of a contrastive loss function, which compares two samples and minimizes the distance between them if they belong to the same class while maximizing the distance between them if they belong to different classes ~\cite{hadsell2006dimensionality}. When looking into tasks that have been tackled using distributed Federated Contrastive Learning (FCL), there are a few applications like image classification~\cite{lewis2019unsupervised}, language modeling ~\cite{tian2020contrastive}, speech recognition~\cite{badziahin2019simultaneous}, and for obtaining useful representations of data for various tasks~\cite{gidaris2018unsupervised}. Due to the increase of data at an unprecedented rate, novel approaches have moved into distributed architectures of contrastive learning. \par
When looking into some of the work that incorporates Contrastive learning successfully, FCL ~\cite{wu2022distributed} integrates FL with Contrastive learning for medical image classifications. FCL trains the model using datasets from multiple parties in a decentralized manner. The parties collaborate to learn a common model but keep their data private. The model improves its performance through Contrastive loss, which minimizes the distance between similar samples and maximizes the distance between dissimilar ones. The goal is to improve accuracy by exchanging encoded features and creating a more diverse dataset. To enhance the model's effectiveness, each client is equipped with remote and local features and performs individual Contrastive learning. The resulting representations are shared through FL to obtain a collective prediction. Overall, in \emph{distributed contrastive learning scape}, there are not many works published due to their complexity and novelty. However, we identify the paper~\cite{wu2022distributed} as a SOTA approach to this topic.

\subsubsection{Knowledge Component in DCL}\hfill\\
In a distributed setting, the \emph{knowledge} of distributed Contrastive learning (DCL) consists of continuous predictive and pretext \emph{logits} that are shared among the model participants. These logits are used to predict one part of an input from another part of the same input, which is the typical working scheme of contrastive learning. Like centralized learning, distributed learning also has two main stages, supervised tasks and knowledge transfer tasks ~\cite{wu2022distributed}. At the pretext or the supervised stage, the model will be learning the data representations of unlabeled data, and the \emph{knowledge} that we can extract at this stage is the \emph{intermediate data representation}. Then this learned representation will be transferred into the global model to make predictions ~\cite{noroozi2018boosting}. If there are several target models attached to the supervised models, the \emph{knowledge} that would be shared will be \emph{gradients} and \emph{parameters}. During the communication phase of distributed transformers, there are a few additional components that may be transferred. These include the model parameters, activations, and other intermediate values that are necessary for calculating the gradients of the model parameters.

\subsection{Attacks and Vulnerabilities in Distributed Semi-Supervised Learning}

The separate entities involved in distributed semi-supervised learning make distributed semi-supervised learning vulnerable to attacks. It is worth noting that semi-supervised learning relies on a small number of labeled datasets to predict a large number of unlabeled datasets. This can make it easier for attackers to target the model and the communication between the entities. In general, the predictions made on the unlabeled dataset rely heavily on the small amount of labeled data. The \emph{knowledge} transfer at the supervised stage consists of \emph{data representations}. As these data representations can be intercepted easily, adversaries can manipulate them, leading to poisoning attacks. The authors of ~\cite{carlini2021poisoning} exploited this vulnerability practically by placing malicious unlabeled data samples totaling only one percent of the entire data set. In the instances where the \emph{data representations} are being shared as \emph{graph embeddings}, the sharing phase consists of unexpected privacy risks ~\cite{zhang2022inference}, such as graph reconstruction, which will allow the attackers to generate graphs that are similar to the structural statistics of the target graph. In the second stage of knowledge sharing in DSSL, we examine the probability that the subgraph contained in the target graph will lead to model inference attacks. We focus on \emph{gradients} and \emph{parameter distributions} that will be transferred when there are two or more target models contributing to a global model. We have found that in these cases, the knowledge components are susceptible to vulnerabilities such as gradient leakage~\cite{tolpegin2020data} and parameter inference~\cite{mcmahan2017communication}, which could compromise the knowledge and exploit the model.\par
The knowledge categories that are mainly affected due to the attacks against DSSML are \emph{neural network update information} and \emph{neural network parameter information}. This clearly highlights the security threat posed by the inconsistencies in labeled data that may be corrupted during the communication process.

\subsection{Defences in Distributed Semi-Supervised learning} 

Semi-supervised learning combines both supervised and unsupervised techniques, so it exhibits characteristics of both. As we have discussed before, when proposing defensive mechanisms, it is important to consider the components that make up the semi-supervised model must work together properly. Incorporating defensive mechanisms for the knowledge components could disrupt the model's functioning or hinder proper collaboration.  When discussing defenses against intercepting \emph{data representation}, we should focus on preserving the privacy and integrity of data states. Bad Data Detection ~\cite{ansari2018graph} is a mechanism that can ensure the \emph{data representations} are accurate with the estimated state metrics using  weighted Steiner tree algorithms ~\cite{klein1995nearly} at the most optimum nodes. By implementing such defense measures, it is possible to protect the privacy of the data used in a DSSL setting. The vulnerabilities we identified include common exploitations resulting from \emph{gradient} and \emph{parameter} leakage. As we have discussed these components in detail in earlier sections, we can use defense strategies that apply to \emph{gradients} and \emph{parameter distributions}.

\section{KNOWLEDGE SHARING IN DISTRIBUTED DEEP REINFORCEMENT LEARNING}
\label{sec:know_DDRL}
Reinforcement Learning has become an essential learning paradigm in AI research. This learning mechanism is crucial for making reliable predictions in more complex and dynamic environments ~\cite{kapturowski2018recurrent}. Since it makes predictions on the dynamic environments, the intelligent agents associated with the learning need to take actions based on the environment in order to maximize the notion of rewarding desired behaviors. The reinforcement learning mechanism is mainly based on making decisions sequentially, and its decisions are dependent on the outputs of the previous input. \emph{Deep Reinforcement Learning} extended to a distributed setting to make the model operate the way it desired. All the components that need to make a \emph{culminative reward} should be shared among the participants alongside the components that are essential for the parallelization of the distributed nodes.

\subsection{Distributed DDRL}
\begin{figure}[t!]
    \centering 
    \includegraphics[width=0.75\linewidth]{ {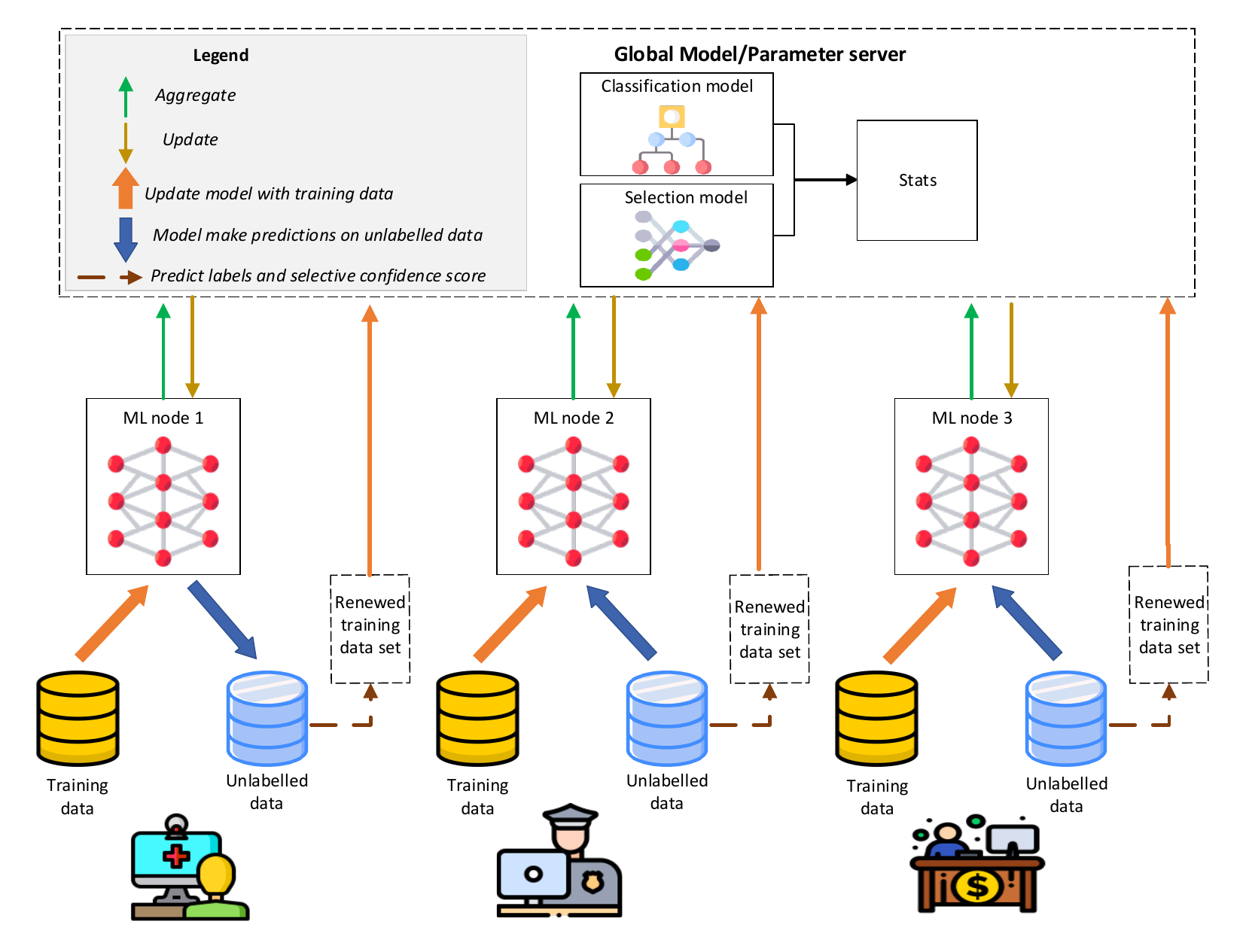} }
    \caption{Knowledge sharing in Distributed Deep Reinforcement Learning.} 
	\label{Fig: Knowledge sharing in DDRL}
	\vspace{-1mm}
\end{figure}

When general neural networks are combined with a  reinforcement learning framework, it is known as \emph{Deep Reinforcement Learning}. This deep reinforcement learning framework efficiently achieves its goals by incorporating techniques such as optimization, function approximation, mapping, and rewards that enhance the performance of the model. This method combines various techniques to make it easier for users to reach their objectives without significant effort. Deep reinforcement learning algorithms also utilize algorithms that identify the most efficient path to achieving pre-defined goals in order to facilitate their learning. When this mechanism applies in a distributed setting, it is known as \emph{Distributed reinforcement learning}. Like most of the deep learning architectures we have discussed before, the distributed deep reinforcement learning methods are based on parallelized training and on Stochastic Gradient Descent~\cite{kovachki2019ensemble}. The general deep reinforcement learning architecture includes an execution environment, actor, and learner. When this architecture, as illustrated in Figure \ref{Fig: Knowledge sharing in DDRL}, is placed in a distributed domain, communication overhead can cause bottlenecks during the \emph{knowledge sharing} phase.
\subsubsection{The Knowledge Component}\hfill\\
The current distributed deep reinforcement architectures are getting improved and becoming more effective over time. The best architecture that demonstrates the working scheme of Distributed Deep Reinforcement Learning (DDRL) can be identified as the GORILA architecture~\cite{nair2015massively}. In GORILA,  there are four main parts which are the parameter server, the worker, the learner, and the replay buffer. This architecture is mainly dependent on the Deep Q-Network (DQN) algorithm. The DQN algorithm is a neural network used for training agents to perform reinforcement learning tasks~\cite{mnih2015human} and makes multiple workers and learners run within a single device and stores Q-networks which is  used to estimate the value of taking a particular action in a given state or the expected return of taking that action. The fundamental approach of this architecture is collaborating with these three entities to optimize the Q-network, which will be later used in the inference stage to make a collective prediction. Distributed Proximal Policy Optimization (PPO) ~\cite{heess2017emergence} is a machine learning technique that is used to train models in a distributed setting. It is an improvement on the traditional PPO algorithm, which belongs to the class of policy gradient algorithms. In distributed PPO, multiple agents or workers are utilized to collect data and perform computations concurrently, allowing for more efficient and resource-saving model training.  Then improving some drawbacks of GORILA~\cite{nair2015massively}, such as correlating and high computational overhead, the architecture of the Actor-Critic ~\cite{mnih2016asynchronous} facilitates DDRL by making multiple agents run asynchronously in parallel to generate data.\par
IMPALA ~\cite{espeholt2018impala} is also a novel DDRL approach similar to the Actor-Critic ~\cite{mnih2016asynchronous} framework, but instead of calculating gradients, it only produces the trajectories of experience that include active state, current state, reward, and memory and communicates those in the global model where it will compute the gradients and optimize both policy and value functions. Ape-X~\cite{horgan2018distributed} and R2D2~\cite{mnih2016asynchronous} are some approaches that implement the distribution of the participants via Recurrent Replay Distributed Reinforcement Learning. Considering the reason this method is unique from the approaches discussed earlier, it aims to reduce variance and accelerate convergence and introduces a gradient prioritization scheme that resolves the gradient overfitting issue that comes with disregarding older experiences with high priorities.  Another novel approach for DDRL is the mechanism called SEED RL~\cite{kapturowski2018recurrent} is a framework that scales up reinforcement learning methods to large environments where large amounts of data are present. This mechanism allows the scaling of larger environments and improves the sample efficiency of the learning process using a distributed architecture, with multiple actors and learners operating in parallel. Also, this mechanism addresses some issues associated with IMPALA~\cite{espeholt2018impala} architecture which bears similar characteristics to this architecture.

\subsection{The attacks and vulnerabilities on DDRL}
When looking into the \emph{knowledge sharing} of DDRL, it can be identified there are several \emph{knowledge components} that are being exploited in modern cyber-attacks. by going through an extensive amount of literature, we present. Table \ref{tab:Knowledge and Vulnerabilities of DDRL} demonstrates those vulnerabilities along with the attacks that exploit them.
\begin{table}
\footnotesize
\begin{longtable}[h!]{|>{\hspace{0pt}}m{0.05\linewidth}|>{\hspace{0pt}}m{0.296\linewidth}|>{\hspace{0pt}}m{0.115\linewidth}|>{\hspace{0pt}}m{0.362\linewidth}|}
\caption{Vulnerabilities in DTL} \label{tab:Knowledge and Vulnerabilities of DDRL}\\ 
\hline
ref & Share point & Knowledge Category & Vulnerability~ \endfirsthead 
\hline
\cite{mnih2016asynchronous} & Policy weights\par{}Perturb weights & Weights & Minimal adversarial perturbations easily fool deep policies.\par{}Policy’s value functions can guide identify when~to inject perturbations.~ ~ ~~ \\ 
\hline
\cite{nair2015massively} & A tuple consisting of a reward, \par{}a new observation, and an end-of-episode indicator.~ ~ ~~ & Reward Memory & Vulnerable to adaptive attacks based on constant estimation\par{}The attacker performs the constant attack based on the estimated upper bounds \\ 
\hline
\cite{kapturowski2018recurrent} & Joint policy that concatenates all the local policies, and\par{}joint state-action trajectories, joint action~ ~ ~~ & Reward\par{}Control Parameter & Dishonest recommendations\par{}The skewed distribution of flows is easy to be compromised by reconnaissance,eavesdrop, and DDOS~ ~ ~~ \\ 
\hline
\cite{espeholt2018impala} & Gradients\par{}Parameters(state,reward,experience) & Gradients\par{}Parameter distribution & Vulnerable to gradient leakage attacks~ ~ ~~ \\ 
\hline
\cite{heess2017emergence} & state-action pair\par{}approximations of the so-called Qmatrices valued\par{}the process that serves as an approximation of the desired value~ ~ ~ & Cell state & By sitting between the agent and the environment, the attacker can monitor \par{}the state, the actions of the agent and the reward signals from the environment.~ ~ ~~ \\ 
\hline
\cite{horgan2018distributed} & Gradients\par{}Aggregated gradients\par{}Updated weights~ ~ ~~ & Gradients\par{}Weights & Gradient disaggregation.\par{}Recursive reconstruction~ ~ ~~ \\
\hline
\end{longtable}
\end{table}
Due to the architectural similarities, the vulnerabilities of GORILLA ~\cite{nair2015massively}, Distributed PPO ~\cite{heess2017emergence} and Asynchronous advantage actor-critic frameworks  were very similar. They demonstrate tendencies for gradient and inference leakage and the ability to infer with the model. Although we have discussed these vulnerabilities when discussing different architectures, knowledge components such as \emph{Experience} and \emph{Update coordinator} stand out the most since they are unique to Reinforcement learning architectures. Sharing the \emph{experience} as the knowledge component will introduce the vulnerability of manipulation by interfering and changing the components in the experience, like the state of the learning iteration, and the reward that will be allocated will corrupt the memory, which will eventually lead to attacks like model poisoning attacks.\par 
When moving into the vulnerabilities that were recognized through IMPALA~\cite{espeholt2018impala} and SEED RL ~\cite{espeholt2019seed} architectures, the main vulnerability that can be identified as \emph{policy}, which is a component that is transferred among participants from the global model. The vulnerabilities associated with these \emph{policy} components can be identified as adversarial attack~\cite{ilahi2021challenges}, overfitting ~\cite{song2019observational}, biasness~\cite{swazinna2021overcoming} and explainability~\cite{heuillet2021explainability}. If an attack intercepts with this policy component, they can define a new meaning for Markov's decision process and can directly comply with the model. So this will eventually lead to model inference or feature inference attacks. The most eye-opening exploitation against Markov's decision processes is by intercepting. With this decision-processing component, the attackers can identify the model's differential threshold. If someone can tailor an attack by looking into Markov's decision process, they can mitigate defensive responses in the attack.\par
Given the complexities in distributed reinforcement learning (DDRL) architectures, it is evident that all knowledge categories within DRL are susceptible to various types of attacks. While the \emph{neural network reinforcement information} category contains the most targeted knowledge components, it is also the category that showcases the key characteristics of reinforcement learning.

\subsection{Defences on DDRL}
The defenses against the attacks on distributed deep reinforcement learning should arise from the reinforcement learning architecture itself ~\cite{shi2021dares}. The model participants in the reinforcement learning process may be biased or dependent on the global parameter server due to the incentive process that drives the reinforcement learning process. Due to its bias and reliance, the parameter server will store user data and model parameters permanently. The global parameter server in the model will have characteristics of a centralized model, which can make it vulnerable to common attacks on centralized learning due to the incentive process that drives the reinforcement learning process. When implementing defensive strategies against attacks on DDRL, we should consider both the architectural integrity of the reinforcement model and its utility.\par
There have been some recent developments in the industry that are used to address these issues. DARES ~\cite{shi2021dares} is one of these strategies. DARES~\cite{shi2021dares} is a system that integrates an asynchronous advantage Actor-Critic model (A3C) and FL to facilitate the device data processing. Using this method, the user is able to process their data locally without collaborating with a third party. To facilitate this working scheme,  DARES uses a local recommendation model trained locally on the user devices using their interaction, and the updates computed locally will be sent to a global recommendation model that is trained on a central server. Another approach is the use of differential privacy to share the global model or the teacher's preliminaries among its students, Gohari et al.~\cite{gohari2021privacy} proposed this defensive strategy which mainly aims to preserve the privacy of the teacher model's demonstrations which will be sent into the student model to expedite their training. This mechanism ~\cite{gohari2021privacy} facilitates the learner's capability of collaborating with the collective learning process despite the perturbations induced by the privacy mechanism. DRL is being used increasingly in modern vehicular systems, so containing the privacy of these vehicular units (VU) is essential to maintain roadside safety and user data. To address this issue, ~\cite{wei2022privacy}  presents a data sanitation model that hides sensitive user information at the vehicular nodes. This mechanism only intercepts with fewer parameters at the local model and hides sensitive information using perturbation techniques by keeping the utility.

\section{LIMITATIONS OF CURRENT KNOWLEDGE SHARING SCHEMES AND FUTURE DIRECTIONS}
\label{sec:limitations_and_future_directions}

Based on the above discussions, it has become apparent that current knowledge-sharing schemes are limited in several ways. As such, this study aims to identify these limitations and provide a roadmap for future developments in this area. Through our analysis, we have highlighted several areas where current knowledge-sharing schemes fall short, including issues related to utility and privacy. These limitations underscore the importance of further research and innovation in knowledge sharing.

\subsection{Preserving Privacy While Keeping the Utility of the Initial Model as Maximum as Possible}
Under the privacy preservation of the knowledge components, most of the defensive mechanisms are associated with encryption, obfuscation, and aggregation techniques. When looking into these mechanisms individually, we note that even though these are effective against some exploitations, they introduce considerable computation overhead to the distributed architecture. General encryption mechanisms bring unbearable communication overhead among the participants because the distributed parties need to encrypt and decrypt the knowledge components ~\cite{elmisery2010privacy}. In order to address this, homomorphic encryption can be introduced ~\cite{zhang2020batchcrypt}. This mechanism enables the distributed parties to perform computations on the encrypted messages without decryption in the training process. Even though homomorphic encryption methods also bring some amount of communication overhead, it has proven their effectiveness through modern applications. Obfuscation-based privacy preservation is primarily achieved through differential privacy but has limited flexibility when handling heterogeneous data. Future research should aim to develop adaptive differential privacy mechanisms that can effectively handle large heterogeneous data silos. 

\subsection{Privacy Preserving Communication Medium Between Distributed Entities}
Although we can use various defensive mechanisms to facilitate \emph{Private knowledge sharing} with optimum utility, if the communication channel between the nodes is compromised, that can compromise the privacy of the entire distributed learning process. So ensuring the privacy of the communication medium among the distributed parties is a must. Generally, the communication between the ML nodes will be initiated through TCP packets. Although this method is proven reliable, the data in a TCP packet can be obtained by dropping the packet or connecting to the network as a third party.  So it is necessary developments should be made in this area. The current developments are mainly based on encryption ~\cite{ccavucsouglu2016novel}, which introduced a certain amount of utility loss, as we have investigated previously. As a solution for this would like to propose developing secure communication architecture that guarantees private knowledge sharing. When considering the privacy-preserving communication medium between distributed entities, we can also consider encryption mechanisms involving HTTP Secure (HTTP) or Transport Layer Security (TLS). HTTPS helps to prevent man-in-the-middle attacks by encrypting the communication between the server and the client, making it difficult for an attacker to intercept and alter the data being transmitted. In contrast, TLS is a protocol for creating secure communication channels between computers on a network. It replaces the older SSL protocol and is used to secure various types of network communication. TLS establishes a secure connection between two devices, such as a client and a server, by negotiating security parameters and exchanging keys. Once the connection is established, the devices can communicate securely over the network.

\subsection{Balancing the Utility of the Model and Privacy of Knowledge Components}
From our survey, we found that the techniques used to preserve knowledge privacy can negatively impact the performance and interpretability of the original AI model and are computationally costly to implement ~\cite{arachchige2019local}. Due to this reason, the question arises whether \emph{"Is it worthwhile to compromise the model's prediction  over privacy-preserved knowledge?"} Striking the right balance between privacy and accuracy in AI models is a complex and nuanced task that depends on the specific context and use case. It is essential to consider the needs and expectations of the stakeholders involved, including legal and ethical considerations. Additionally, it is important to involve data protection and privacy experts in designing the AI model. Furthermore, it is not just the amount of data that affects the accuracy of the model but also factors such as the quality of the data, the model architecture, and the training process. To make an informed decision, it is crucial to thoroughly understand the problem at hand and evaluate all options and trade-offs before proceeding.\par
Achieving a balance between privacy and accuracy in AI models is a multifaceted challenge. Some methods to address this issue may involve:
\begin{itemize}
    \item \emph{Conducting a risk assessment} ~\cite{hogan2021ethics} is crucial in finding the balance between privacy and accuracy in AI models. This process involves identifying the types of personal data that will be used, assessing the potential risks associated with using that data and determining the measures that will be taken to mitigate those risks. 
    \item  \emph{Continual monitoring and evaluation} ~\cite{zhao2017machine} of the AI model's performance and the implementation of privacy measures are necessary to identify and address any issues or concerns. This helps to ensure that the AI model is operating as intended and that the privacy measures  are adequate and effective. 
    \item  \emph{Anonymization and de-identification techniques} ~\cite{andrew2019privacy} can be applied to remove or mask personal identifiers from the data, making it difficult or impossible to re-identify an individual. This can help to protect personal data while preventing knowledge components from being reconstructed by the attacker.
\end{itemize}
Finding the balance between privacy and accuracy is an iterative process that may require experimentation and adjustments. The best approach will depend on the specific use case of the AI implementations and the architectural needs for the collaboration involved.


\section{CONCLUSION}
\label{sec:conclusion}


This paper has offered a comprehensive analysis of the diverse types of \emph{knowledge} that can be exchanged during collaborative AI/ML operations among distributed entities. In addition, it has examined the potential vulnerabilities of these knowledge components and how cyber attackers could exploit them to compromise distributed systems. The study has also presented the existing defensive measures for mitigating these identified vulnerabilities. Furthermore, this article has underscored the shortcomings of current knowledge-sharing techniques and emphasized the need for future improvements to enhance the utility and privacy of \emph{knowledge sharing} in distributed learning. As such, it has provided valuable insights into the current state of knowledge sharing in collaborative AI/ML operations and highlighted the importance of safeguarding distributed systems against cyber threats.

\bibliographystyle{unsrt}
\bibliography{main}

\end{document}